\documentclass[10pt,english,aps,pra,longbibliography,twocolumn]{revtex4-1}

\usepackage[T1]{fontenc}
\usepackage{graphicx}% Include figure files
\usepackage{bbm}% bold math
\usepackage[usenames,dvipsnames]{xcolor}

\usepackage{stmaryrd}
\usepackage{tabularx}
\usepackage{float}
\usepackage{array}
\newcommand{\rvline}{\hspace*{-\arraycolsep}\vline\hspace*{-\arraycolsep}}

\usepackage{amsmath} 
\usepackage[colorlinks=true,citecolor=blue,linkcolor=RubineRed,urlcolor=blue]{hyperref}
\usepackage[normalem]{ulem} %normalem is useful to remove the underlining of the title in the bibliography for books!
\usepackage{MnSymbol}
\providecommand{\llangle}{\langle\!\langle}
\providecommand{\rrangle}{\rangle\!\rangle}

\begin{document}

\title{Third-quantized master equations as a classical Ornstein-Uhlenbeck process}

\author{L\'eonce Dupays   \href{https://orcid.org/0000-0002-3450-1861}{\includegraphics[scale=0.05]{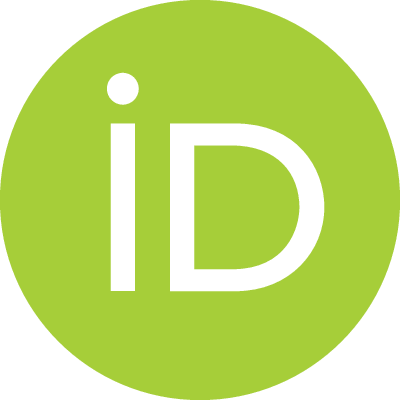}}}
\email{leonce.dupays@gmail.com}
%\affiliation{Department of Mathematics, King’s College London, Strand, London WC2R 2LS, UK}
\affiliation{Department  of  Physics  and  Materials  Science,  University  of  Luxembourg,  L-1511  Luxembourg,  Luxembourg}
\begin{abstract}
Third quantization is used in open quantum systems to construct a superoperator basis in which quadratic Lindbladians can be turned into a normal form. From it follows the spectral properties of the Lindbladian, including eigenvalues and eigenvectors. However, the connection between third quantization and the semiclassical representations usually employed to obtain the dynamics of open quantum systems remains opaque. We introduce a new basis for third quantization that bridges this gap between third quantization and the $Q$ representation by projecting the master equation onto a superoperator coherent-state basis. The equation of motion reduces to a multidimensional complex Ornstein-Uhlenbeck process. \\

\end{abstract}
\maketitle
\section{Introduction}
Open quantum systems are one of the pillars of contemporary physics, ranging from the description of light-matter interaction \cite{Rivas_2012} and the theory of quantum measurement \cite{Elouard2017} to topological phases of matter \cite{Gong2018} and quantum algorithms \cite{Miessen2023,Delgado_Granados2025}. The equation of motion of a Markovian open quantum system is well described by the Lindblad master equation \cite{Lindblad1976}, whose solution is the reduced density matrix. To solve its dynamics, or to compute expectation values of an observable, a wide range of techniques have been developed. A general method consists of projecting the dynamics onto the coherent-state basis \cite{Glauber1963} and studying the evolution of quasiprobabilities, such as the $P$ and $Q$ representation. For quadratic systems, another approach involves introducing a superoperator basis, known as third quantization \cite{Prosen_2008,Prosen_2010}, and to bring the Lindbladian to a normal form. Furthermore, the dynamics of Gaussian states is known from their covariance matrix dynamics \cite{Weedbrook2012}, but third quantization proves useful in characterizing the steady state for any initial state. However, the connection between the normal form approach and the classical representation remains elusive. Recently, the authors of \cite{McDonald2023} showed that by choosing an appropriate third-quantization basis, one can easily determine the equation of motion for the Wigner function. In this article we demonstrate that another choice of third-quantization basis leads us to consider the $Q$ representation and its characteristic function as ``wave functions'' of the density matrix, which allows us to determine the equations of the quantum-classical correspondence for the $Q$ representation. 

We further demonstrate that the equation of motion followed by the $Q$ representation is the one of an Ornstein-Uhlenbeck process. We provide an explicit mapping between the structure of the quadratic master equation, defined by its Hamiltonian, loss, pumping, and coherence matrices, and the drift and diffusion matrices of the Ornstein-Uhlenbeck process. These matrices encode the thermodynamic properties of the system. The $Q$ representation is particularly well suited for characterizing the entropy, as it always remains positive, unlike other quasiprobability distributions, which justifies the relevance of the alternative third-quantized basis. The entropy associated with the $Q$ representation is known as the Wehrl entropy \cite{Wehrl1979}. 

To achieve this, we primarily focus on how a real-space third-quantization basis allows us to obtain the normal form of the multidimensional real Ornstein-Uhlenbeck process. From this normal form, we can determine the eigenvalues and eigenvectors of the multidimensional Ornstein-Uhlenbeck process. This result has potential applications in classical physics and stochastic processes. For example, studying the eigenstates of the Ornstein-Uhlenbeck process has been shown to aid in understanding the asymptotic phases of stochastic oscillators \cite{Thomas2014,Thomas2021} and limit-cycle behavior in phase space \cite{Dutta2024}. 

By interpreting the coherent-state basis as analogous to the position basis in the complex plane, we show that the quadratic bosonic Lindblad equations of motion form a complex multidimensional Ornstein-Uhlenbeck process. We obtain the normal form of the quadratic bosonic master equation in a manner similar to that used for the Ornstein-Uhlenbeck process.

{
We close by demonstrating that the third-quantized basis introduced by Prosen and Seligman is useful to determine the equations of motion in the $P$ representation. As an illustrative example, we study a dissipative harmonic oscillator in a squeezed thermal bath and provide the equations of motion in the three third-quantized bases, along with their associated quasiprobability distribution representation. 
}
The paper is structured as follows. In Sec.~\ref{normal_ornstein_uhlenbeck} we determine the normal form of the classical Ornstein-Uhlenbeck process and compute its right and left eigenvectors. In Sec.~\ref{third_quantized_basis} we introduce an alternative third-quantized basis. In Sec.~\ref{superoperator_coherent_state} we introduce the superoperator coherent state basis and its Fourier conjugate. In Sec.~\ref{Q_representation} we project the third-quantized master equation onto the superoperator coherent state basis and obtain the equation of motion for the $Q$ representation in the form of an Ornstein-Uhlenbeck process. In Sec.~\ref{eigenvectors_Q_representation} we establish the eigenvectors of the $Q$ representation. In Sec.~\ref{evolution_covariance} we present the equations of evolution for the covariance matrix. In Sec.~\ref{evolution_Q_representation} we focus on the time evolution of the $Q$ representation. In Sec.~\ref{prosen_seligman} we show how the $P$ representation can be obtained using the Prosen and Seligman basis. In Sec.~\ref{example} we illustrate our formalism on the example of a squeezed thermal bath, and derive the equations of motion for the different third quantized bases, and their associated quasi-probability distribution. Finally, in Sec.~\ref{conclusion} we provide a conclusion and outlook.

\section{ Normal form of the Ornstein-Uhlenbeck process \label{normal_ornstein_uhlenbeck}} 
As a preliminary step, let us derive the normal form of the Ornstein-Uhlenbeck (OU) process. The OU process was originally introduced to describe the velocity of massive particles under friction \cite{Uhlenbeck1930}. However, its applications have largely surpassed its original purpose, ranging from random matrix theory \cite{Dyson1962} and active matter \cite{Martin2021} to networks of noisy synchronized oscillators \cite{Elowitz2000,Katoh2020}. Its spectral properties illuminate the dynamics of the system. Eigenvalues provide information on the rate of convergence to the nonequilibrium steady state or oscillatory behavior. Eigenstates help determine the dynamical evolution for an arbitrary initial state. Additionally, it is possible to solve the dynamics in Fourier space with a Gaussian ansatz \cite{Risken1996,Vatiwutipong2019} for specific initial conditions. The study of the eigenstates also informs on asymptotic phases of stochastic oscillators \cite{Thomas2014,Thomas2021} and on the limit cycle behavior in phase space \cite{Dutta2024}. Previous works that study the eigenstates of the multidimensional OU process are restricted to the case of simultaneously diagonalizable drift and diffusion matrices \cite{Liberzon2000}, the case of a normal drift matrix and a diffusion matrix proportional to identity \cite{Chen2014}, and the case of a diagonalizable drift matrix \cite{Leen2016}. We demonstrate in the following that the OU process can be recast in a normal form and from this build its eigenvectors. Let us consider an $N$-dimensional OU process defined by
\begin{align}
d x_{t}=-\beta x_{t}dt+\sigma dW_{t}.\label{eq:stochastic}
\end{align}
Here $W_{t}$ is an $N$-dimensional Wiener process, and $\beta$ and $\sigma$ are $N\times N$ constant matrices. $dW_{t}=W(t+t')-W(t')$ is the Wiener increment and the Wiener process $W(t)=\int_{0}^{t}\xi_{t'}dt'$, where $\xi_{t}$ is the Gaussian noise with $\langle \xi_{t'}\xi_{t}\rangle=\delta(t-t')$. The Wiener process satisfies the properties of stochastic calculus such that for any function $f$ multiplied by a Wiener process, $\langle f(t)W(t)\rangle=0$. The correlation function is $\langle W(t_{1})W(t_{2})\rangle=2t_{2}$ for $t_{1}\geq t_{2}$ and $\langle W(t_{1})W(t_{2})\rangle=2t_{1}$ otherwise. The OU process relies on additive noise, so that Itô and Stratonovich integrations are equivalent in this case \cite{Risken1996}. The evolution of the probability density function $P(\underline{x},t)$, with the position vector denoted $\underline{x}^{T}=(x_{1}\ x_{2}\ \cdots\ x_{n})$, where the exponent $^{T}$ stands for the transpose, can be described by a Fokker-Planck equation. From the Kramers-Moyal expansion, one can obtain the forward and adjoint backward Fokker-Planck propagators \cite{Risken1996}
\begin{align}
L_{\rm FP}(\underline{x},t)&=-\sum_{i=1}^{N}\partial_{x_{i}}D_{i}(\underline{x},t)+\sum_{i,j=1}^{N}\frac{\partial^{2}}{\partial_{x_{i}}\partial_{x_{j}}}D_{ij}(\underline{x},t),\\
L^{+}_{\rm FP}(\underline{x}',t)&=\sum_{i=1}^{N}D_{i}(\underline{x}',t)\partial_{x'_{i}}+\sum_{i,j=1}^{N}D_{ij}(\underline{x}',t)\frac{\partial^{2}}{\partial_{x'_{i}}\partial_{x'_{j}}},\label{eq:backpropagator}
\end{align}
where the symbol $+$ stands for the backward propagator. In particular, for the OU process, $D_{i}(\underline{x},t)=-\sum_{j}\beta_{ij}x_{j}$ and $D_{ij}(\underline{x},t)=D_{ij}$, with $D_{ij}$ being the components of the diffusion matrix $D$. Here, $D={\bf \sigma}{\bf \sigma}^{T}/2$, and $\beta$ is the drift coefficient matrix of components $\beta_{ij}$. Finally, one can write the equation of evolution of the probability distribution for the forward process 
\begin{align}
\partial_{t}P=\sum_{i,j}\beta_{ij}\partial_{x_{i}}(x_{j}P)+\sum_{i,j}D_{ij}\partial_{x_{i}}\partial_{x_{j}}P. \label{OU_process_eq}
\end{align}
To study the spectral decomposition of the OU process, it is convenient to recast the Fokker-Planck equation in a normal form. To this end, we introduce the superoperator basis
\begin{align}
{\bf c}_{j}&=x_{j}\bullet,\ {\bf c}'_{j}=-\partial_{x_{j}}\bullet
\end{align}
that verify the commutation relations $[{\bf c}_{i},{\bf c}'_{j}]=\delta_{ij}\mathbbm{1}\bullet$. These operators play a similar role to position and momentum operators in quantum mechanics. One can then define the vectors $\underline{{\bf c}}^{T}_{j}=({\bf c}_{1}\ \cdots\ {\bf c}_{N})$,  $\underline{{\bf c}}'^{T}_{j}=({\bf c}'_{1}\ \cdots\ {\bf c}'_{N})$ and write the OU process as a quadratic form
\begin{align}
\partial_{t}P=\frac{1}{2}(\underline{{\bf c}}^{T}S_{-}\underline{{\bf c}}+{\rm Tr}\beta)P=\underline{L}P,
\end{align}
with $\underline{{\bf c}}^{T}=(\underline{{\bf c}}^{T}_{j}\ \underline{{\bf c}}'^{T}_{j})$ and 
\begin{align}
 S_{\mp}&=\begin{pmatrix}
  0 & \rvline & \mp\beta^{T} \\
\hline
 \mp \beta & \rvline &
  2D
\end{pmatrix}\label{eq:matrix_symmetric}.
\end{align}
We note that the full matrix $S_{\mp}$ is symmetric given that $D^{T}=D$. Assuming that the matrix $\beta$ is diagonalizable \cite{Prosen_2010} and writing $\beta=P\Delta P^{-1}$ with $\Delta={\rm diag}(\beta_{1},\cdots,\beta_{N})$, we use the known normal form decomposition \cite{Prosen_2008} to write the Liouvillian $\underline{L}$ in a normal form. The procedure is detailed in Appendix \ref{normal_form_OU} so that finally 
\begin{align}
\underline{L}=-\frac{1}{2}\sum_{i=1}^{N}\beta_{i}(\zeta'_{i}\zeta_{i}+\zeta_{i}\zeta'_{i})+\frac{1}{2}{\rm Tr}\beta,
\end{align}
where the new operators components of the vectors $\underline{\zeta}^{T}=(\zeta_{1} \cdots \zeta_{N})$ and $\underline{\zeta}'^{T}=(\zeta'_{1} \cdots \zeta'_{N})$ are defined from the canonical transformation $(\underline{\zeta},\underline{\zeta}')^{T}=(V\underline{c})^{T}$, and
\begin{align}
V=\begin{pmatrix}
P^{-1}& -P^{-1}\Sigma_{\infty}\\
0 & P^{T}
\end{pmatrix}\label{eq:passage},
\end{align}
and $\Sigma_{\infty}$ is the solution to the Lyapunov equation 
\begin{align}
\beta\Sigma_{\infty}+\Sigma_{\infty}\beta^{T}=2D.\label{eq:Lyapunov}
\end{align}
Then one can write the new operators as
\begin{align}
\zeta_{i}&=\sum_{j}P^{-1}_{ij}x_{j}+\sum_{j}(P^{-1}\Sigma_{\infty})_{ij}\partial_{x_{j}},\label{annihil_ornstein}\\
\zeta'_{i}&=-\sum_{j}P^{T}_{ij}\partial_{x_{j}}. \label{crea_ornstein}
\end{align}
The canonically transformed operators obey the bosonic commutation relations $[\zeta_{i},\zeta'_{j}]=\delta_{ij}$. Finally, the Liouvillian reduces to the normal form
\begin{align}
\underline{L}=-\sum_{i=1}^{N}\beta_{i}\zeta'_{i}\zeta_{i}.\label{Liouvillian_real_ornstein}
\end{align}
One can verify in inserting \eqref{annihil_ornstein} and \eqref{crea_ornstein} in \eqref{Liouvillian_real_ornstein} that the OU process is recovered \eqref{OU_process_eq}. It happens that these operators are annihilation and creation operators for the OU process. Indeed, the right vacuum of the Liouvillian is the nonequilibrium steady state (NESS)
\begin{align}
\underline{L}P_{\rm ness}(\underline{x})=0.
\end{align}
However, this is also the vacuum of the annihilation operator, which can simply be identified as $\zeta_{i}$
\begin{align}
\zeta_{i}P_{\rm ness}(\underline{x})=0.
\end{align}
We note that contrary to the harmonic oscillator normal form, the vacuum energy for the OU process is zero, which is associated to the fact that the equivalent to the vacuum state for open systems is the non-equilibrium steady state. From here on, one can build the eigenvector Fock basis of the Liouvillian
\begin{align}
r_{\underline{\mu}}(\underline{x}) &= \prod_{i}\frac{(\zeta'_{i})^{\mu_{i}}}{\sqrt{\mu_{i}!}}P_{\rm ness}(\underline{x}), 
\end{align}
where $\mu_{i}$ represents the occupation number in the Fock basis for each mode. The eigenoperator of the adjoint Liouvillian $\underline{L}^{+}$ is denoted $l_{\mu}(\underline{x})$. They share the same eigenvalues \cite{Risken1996}, which leads to the spectral decomposition of the Liouvillian
\begin{align}
\underline{L}r_{\underline{\mu}}(\underline{x})&=E_{\underline{\mu}}r_{\underline{\mu}}(\underline{x}),\underline{L}^{+}l_{\underline{\mu}}(\underline{x})=E_{\underline{\mu}} l_{\underline{\mu}}
(\underline{x}),\ E_{\underline{\mu}}=-\sum_{i=1}^{N}\mu_{i}\beta_{i},
\end{align}
where $E_{\underline{\mu}}$ is the eigenvalue of the Liouvillian. The scalar product between the right eigenvector and its adjoint is given by \cite{Risken1996}
\begin{align}
\langle l_{\underline{\mu}},r_{\underline{\nu}}\rangle=\int d^{N}x\  l_{\underline{\mu}}(\underline{x})r_{\underline{\nu}}(\underline{x})=\delta_{\underline{\mu},\underline{\nu}},
\end{align}
where $d^{N}x=dx_{1}\cdots dx_{N}$. The evolution of the probability distribution is known from the spectral decomposition, so that the transition probability density reads \cite{Risken1996}
\begin{align}
P(\underline{x},t|\underline{x}',0)&=e^{\underline{L}t}\delta(\underline{x}-\underline{x}')\nonumber\\
&=\sum_{\underline{\mu}}e^{E_{\underline{\mu}}t}r_{\underline{\mu}}(\underline{x}) l_{\underline{\mu}}(\underline{x}').
\end{align}
The spectrum $E_{\underline{\mu}}$ characterizes the dynamics of the system. If the spectrum is real and strictly negative, $E_{\underline{\mu}}<0$, the system relaxes to a NESS. In the presence of imaginary parts, oscillatory phases appear \cite{Thomas2014,Thomas2021}. Consider the decomposition $E_{\underline{\mu}}=\mathcal{R}e[E_{\underline{\mu}}]+i\mathcal{I}m[E_{\underline{\mu}}]$. The condition $\mathcal{R}e[E_{\underline{\mu}}]\ll|\mathcal{I}m[E_{\underline{\mu}}]|$ guarantees that the relaxation is small enough for the oscillatory phase to alter the dynamics.

By following the same reasoning, one can obtain the normal form decomposition for the adjoint operator. The adjoint propagator flips the sign of the drift matrix $\beta \to -\beta$ as seen from \eqref{eq:backpropagator} so that $\underline{L}^{+}=\frac{1}{2}(\underline{c}^{T}S_{+}\underline{c}+{\rm Tr}\beta)$. By following the normal form decomposition procedure for the backward propagator \eqref{eq:backpropagator}, with $\beta=P\Delta P^{-1}$, one obtains a new passage matrix
\begin{align}
V^{+}&=\begin{pmatrix}
P^{-1} & P^{-1}\Sigma_{\infty}\\
0& P^{T} 
\end{pmatrix}.\label{eq:passage_plus}
\end{align}
Note the sign difference with respect to \eqref{eq:passage}. The ladder operators are given by the transformation $(\underline{\zeta}'^{+},\underline{\zeta}^{+})^{T}=(V^{+}\underline{c})^{T}$ so that
\begin{align}
\zeta'^{+}_{i}&=\sum_{j}P^{-1}_{ij}x_{j}-\sum_{j}(P^{-1}\Sigma_{\infty})_{ij}\partial_{x_{j}},\label{eq_+_crea}\\
\zeta^{+}_{i}&=-\sum_{j}P^{T}_{ij}\partial_{x_{j}}. \label{eq_+_annihil}
\end{align}
that verify the commutation relation $[\zeta'^{+}_{i},\zeta^{+}_{j}]=\delta_{ij}\mathbbm{1}$. Finally, the normal form of the adjoint operator is given by 
\begin{align}
\underline{L}^{+}&=\frac{1}{2}\sum_{i=1}^{N}\beta_{i}(\zeta'^{+}_{i}\zeta^{+}_{i}+\zeta^{+}_{i}\zeta'^{+}_{i})+\frac{1}{2}{\rm Tr}\beta. \label{normal_L+}
\end{align}
Inserting \eqref{eq_+_annihil} and \eqref{eq_+_crea} in the normal form decomposition \eqref{normal_L+}, one recovers the adjoint Liouvillian given by \eqref{eq:backpropagator}. Furthermore, the vaccum state of the adjoint propagator is the identity operator 
\begin{align}
\underline{L}^{+}\mathbbm{1}=0.
\end{align}
However, the operator that annihilates the vacuum state is the annihilation operator for the Liouvillian, so that the annihilation operator is in fact $-\zeta^{+}_{i}$, with the commutation relation $[-\zeta^{+}_{i},\zeta'^{+}_{j}]=\delta_{ij}\mathbbm{1}$ and, $\forall i$
\begin{align}
-\zeta^{+}_{i}\mathbbm{1}=0.
\end{align}
From here on, one can construct the eigenvector basis
\begin{align}
l_{\underline{\mu}}(\underline{x})&=\prod_{i=1}^{N}\frac{(\zeta'^{+}_{i})^{\mu_{i}}}{\sqrt{\mu_{i}!}}\mathbbm{1}.
\end{align}
At the end, the Liouvillian takes the form 
\begin{align}
\underline{L}^{+}&=-\sum_{i=1}^{N}\beta_{i}\zeta'^{+}_{i}(-\zeta^{+}_{i}).
\end{align}
Let us now determine the form of the eigenstates of the OU process. The covariance matrix of the OU process \cite{Risken1996,Vatiwutipong2019}
\begin{align}
\Sigma_{t}&=\left\langle(\underline{x}_{t}-\langle \underline{x}_{t}\rangle)(\underline{x}_{t}-\langle \underline{x}_{t}\rangle)^{T}\right\rangle.
\end{align}
with $\langle\underline{x}_{s}\rangle=\int d^{N}\underline{x} P(\underline{x},s)\underline{x}$ and, $d^{N}\underline{x}=dx_{1}\cdots dx_{N}$. It is known that it obeys the differential equation \cite{Risken1996}
\begin{align}
\frac{d\Sigma_{t}}{dt}&=-\beta\Sigma_{t}-\Sigma_{t}\beta^{T}+2D.
\end{align}
It can be demonstrated by adopting a Gaussian ansatz for the characteristic function, which is Fourier transform of the transition probability density \cite{Risken1996}. As a consequence, the stationary autocorrelation function obeys the Lyapunov equation \eqref{eq:Lyapunov}. This is a multidimensional extension of the well-known relationship for the one-dimensional $(1D)$ OU process, where the stationary covariance is given by $\Sigma^{1D}_{\infty}=\frac{D^{1D}}{\beta^{1D}}$. Additionally, for the OU process, the nonequilibrium stationary probability distribution can be described in terms of the stationary covariance matrix \cite{Risken1996}
\begin{align}
P_{\rm ness}(\underline{x})=\frac{\exp[-\frac{1}{2}\underline{x}^{T}\Sigma^{-1}_{\infty}\underline{x}]}{(2\pi)^{N/2}({\rm det}\Sigma_{\infty})^{1/2}}. \label{pness}
\end{align}
The stationary covariance matrix, $\Sigma_{\infty}$, is assumed to be invertible. Note that, at infinite time, the mean of the process converges to zero, which is a consequence of the mean-reverting property of the OU process. This mean-reverting characteristic can be generalized to an arbitrary mean by introducing an asymptotic mean parameter in the stochastic equation \eqref{eq:stochastic} \cite{Vatiwutipong2019}. From the knowledge of the NESS, it is possible to construct the eigenvectors. Assuming the whitening decomposition of the inverse covariance matrix $\Sigma^{-1}_{\infty}=W^{T}W$ and the change of variable 
\begin{align}
\underline{x}'&=W\underline{x},\\
\underline{\partial}_{x}&=W^{T}\underline{\partial}_{x'}.
\end{align}
One can find the right eigenstate of the multidimensional OU process 
\small
\begin{align}
r_{\underline{\mu}}(\underline{x})&=P_{\rm ness}(\underline{x})\sum_{\substack{k^{1}_{1}+k^{1}_{2}+\cdots+k^{1}_{N}=\mu_{1},\\
\vdots\\
k^{N}_{1}+k^{N}_{2}+\cdots+k^{N}_{N}=\mu_{N}}}\Big[\prod_{i=1}^{N}\frac{1}{\sqrt{\mu_{i}!}}\binom{\mu_{i}}{k^{i}_{1},k^{i}_{2},\cdots,k^{i}_{N}}\Big]\nonumber\\
&\times\prod_{t=1}^{N}\Big\{\prod_{i=1}^{N}[(WP)^{T}]^{k^{i}_{t}}_{it}\Big\}H_{\sum_{i=1}^{N}k^{i}_{t}}(x'_{t}),
\end{align}
\normalsize
where $H_{n}(x)=(-1)^{n}e^{x^{2}/2}\partial^{n}_{x}e^{-x^{2}/2}$ is the probabilist Hermite polynomial. And the multinomial coefficient
\begin{align}
\binom{\mu_{i}}{k_{1},k_{2},\cdots,k_{N}}=\frac{\mu_{i}!}{k_{1}!k_{2}!\cdots k_{N}!}.
\end{align}In particular, the NESS is recovered for  $\underline{\mu}=\underline{0}$. The left eigenstate is given by 
\begin{align}
l_{\mu}(\underline{x})&=\sum_{\substack{k^{1}_{1}+k^{1}_{2}+\cdots+k^{1}_{N}=\mu_{1},\\
\vdots\\
k^{N}_{1}+k^{N}_{2}+\cdots+k^{N}_{N}=\mu_{N}}}\Big[\prod_{i=1}^{N}\frac{1}{\sqrt{\mu_{i}!}}\binom{\mu_{i}}{k^{i}_{1},k^{i}_{2},\cdots,k^{i}_{N}}\Big]\nonumber\\
&\times\prod_{t=1}^{N}\Big\{\prod_{i=1}^{N}(P^{-1}W^{-1})^{k^{i}_{t}}_{it}\Big\}H_{\sum_{i=1}^{N}k^{i}_{t}}(x'_{t}).
\end{align}
The identity operator is recovered for $\underline{\mu}=0$. The proof for the two formulas is found in Appendix \ref{app:eigenstates}.

\section{Third quantized basis \label{third_quantized_basis}}
To determine a quantum equivalent to the OU process, we replace the probability density function $P(x)$ by the $Q$ representation of the density matrix. The $Q$ representation is defined as $Q(\alpha)=\langle \alpha|\rho|\alpha\rangle/\pi$, where $|\alpha\rangle$ are the coherent states \cite{Glauber1963}. Additionally, we replace the position basis variable $x$ by the coherent state parameter $\alpha$. 

In this context we introduce the Hilbert space of element $|\psi\rangle \in \mathcal{H}$ of $N$ bosons. These elements verify the canonical commutation relations: $[a_{n},a^{\dagger}_{m}]=\delta_{nm}$, where $a_{n}$ and $a^{\dagger}_{m}$ are the annihilation and creation operators, respectively. To proceed further, we introduce the Liouville representation $|A\rho B\rangle =(A\otimes B^{T})|\rho\rrangle$. $|\rho\rrangle $ stands for the vectorized density matrix \cite{Gyamfi_2020}, with the linear map
\begin{eqnarray}
\Omega\left[|\Psi\rangle \langle \phi |\right] \rightarrow  |\Psi \rangle \otimes |\phi \rangle^{*} \equiv |\Psi,\phi \rrangle , \quad \forall~ \Psi,\phi \in \mathcal{H},
\end{eqnarray}
where $^{*}$ denotes the complex conjugation, $\otimes$ the tensorial product, and $\mathcal{H}$ the system Hilbert space. This corresponds to rearranging the rows of the density matrix in columns \cite{Gyamfi_2020}. The scalar product in the vectorized representation is defined as $\llangle X|Y\rrangle\equiv{\rm Tr}[X^{\dagger}Y]$.

To ensure the quantum systems exhibit properties analogous to the classical OU process, we introduce the following superoperators
\begin{align}
{\bf c}'_{0,n}|\bullet\rrangle&=|\bullet a_{n}\rrangle,\quad{\bf c}_{0,n}|\bullet\rrangle=|-[a^{\dagger}_{n},\bullet]\rrangle,\\
{\bf c}'_{1,n}|\bullet\rrangle&=|a^{\dagger}_{n}\bullet\rrangle,\quad{\bf c}_{1,n}|\bullet\rrangle=|[a_{n},\bullet]\rrangle.
\end{align}
They act on the doubled Hilbert space of element $|\phi, \psi\rrangle \in \mathcal{H}\otimes \mathcal{H}$ and verify the commutation relations $[{\bf c}_{0,n},{\bf c}'_{0,m}]=[{\bf c}_{1,n},{\bf c}'_{1,m}]=\delta_{nm}\mathbbm{1}\bullet$ and $[{\bf c}_{i,n},{\bf c}_{j,m}]=[{\bf c}'_{i,n},{\bf c}'_{j,m}]=0$ for $i\neq j$. Note that this basis is the complex conjugated of the one introduced by Prosen and Seligman \cite{Prosen_2010}. For clarity, we detail the vectorization procedure for the bosonic operator ${\bf c}_{1}$, and one can write its action on the density matrix
\begin{align}
{\bf c}_{1}|\rho\rrangle&=|a\rho-\rho a\rrangle =\left(a \otimes \mathbbm{1}-\mathbbm{1}\otimes (a )^{T}\right)|\rho\rrangle.
\end{align}
The annihilation operator can be represented by the infinite dimensional real matrix in the Fock basis
\begin{align}
a&=\begin{pmatrix}
0 & \sqrt{1} & 0 & 0 & \dots & 0 & \dots \\
0 & 0 & \sqrt{2} & 0 & \dots & 0 & \dots \\
0 & 0 & 0 & \sqrt{3} & \dots & 0 & \dots \\
0 & 0 & 0 & 0 & \ddots & \vdots & \dots \\
\vdots & \vdots & \vdots & \vdots & \ddots & \sqrt{n} & \dots \\
0 & 0 & 0 & 0 & \dots & 0 & \ddots \\
\vdots & \vdots & \vdots & \vdots & \vdots & \vdots & \ddots \end{pmatrix},\label{rep_annihilation}
\end{align}
so that from a matrix point of view $a^{T}=a^{\dagger}$. In the end of the day, the operator ${\bf c}_{1}$ is represented by the tensorial product matrix $a\otimes \mathbbm{1}-\mathbbm{1}\otimes a^{T}$.  As a further example, the full loss dissipator is treated in Appendix~\ref{vectorization}. Superoperators ${\bf c}_{0,n}$ and ${\bf c}_{1,n}$ annihilate the vacuum operator, which is the vectorized identity $|\mathbbm{1}\rrangle$ 
\begin{align}
{\bf c}_{0,n}|\mathbbm{1}\rrangle={\bf c}_{1,n}|\mathbbm{1}\rrangle=0,
\end{align}
and ${\bf c}'_{0,n}$, ${\bf c}'_{1,n}$ left-annihilate the vectorized vacuum state $\llangle 0|\bullet\rrangle=\langle 0 |\bullet |0\rangle$
\begin{align}
\llangle 0|{\bf c}'_{0,n}=\llangle 0|{\bf c}'_{1,n}=0.
\end{align}
As a consequence, one can define the left and right eigenoperators, that form a bi-orthogonal basis, so-called third-quantized basis
\begin{align}
|p_{\underline{\mu},\underline{\nu}}\rrangle&=\bigotimes_{n}\frac{({\bf c}'_{0,n})^{\mu_{n}}({\bf c}'_{1,n})^{\nu_{n}}}{\sqrt{\mu_{n}!\nu_{n}!}}|\mathbbm{1}\rrangle,\label{eq:right_basis}\\
\llangle q_{\underline{\mu},\underline{\nu}}|&=\llangle 0|\bigotimes_{n}\frac{({\bf c}_{0,n})^{\mu_{n}}({\bf c}_{1,n})^{\nu_{n}}}{\sqrt{\mu_{n}!\nu_{n}!}}.\label{eq:left_basis}
\end{align}
They verify the orthogonality condition $\llangle p_{\underline{\mu},\underline{\nu}}|q_{\underline{\mu}',\underline{\nu}'}\rrangle=\delta_{\underline{\mu},\underline{\mu}'}\delta_{\underline{\nu},\underline{\nu}'}$. The superoperators play the role of ladder operators with well-defined eigenvalues. Note that the choice of the third-quantized basis is arbitrary and other choices are possible \cite{Prosen_2008,Barthel2022,McDonald2023}; however, we justify the choice of this basis to perform the quantum-classical correspondence. Note that at this stage, these eigenfunctions are merely mathematical objects and do not have a clear physical meaning.

\section{Coherent states of the third quantized basis \label{superoperator_coherent_state}}
We just introduced a superoperator Fock basis. In analogy to the coherent states of the Fock basis \cite{Glauber1963}, one can introduce a superoperator coherent basis. Let us introduce the left superoperator coherent state and the right antinormally ordered operator 
\begin{align}
\llangle \underline{\alpha}|\bullet\rrangle&={\rm Tr}[|\underline{\alpha}\rangle\langle \underline{\alpha}|\bullet]=\langle \underline{\alpha}|\bullet|\underline{\alpha}\rangle,\\
|\underline{\eta}_{a}\rrangle&=|\bigotimes_{n}\eta_{a,n}\rrangle,
\end{align}
with $\eta_{a,n}=e^{\eta^{*}_{n}a_{n}}e^{-{\eta}_{n}a^{\dagger}_{n}}$ and the tensorial product of coherent states $|\underline{\alpha}\rangle=\otimes_{n=1}^{N}|\alpha_{n}\rangle$. They are respectively left and right
 eigenvectors of the superoperator basis in a similar way to the coherent states with the annihilation operator $a|\alpha\rangle=\alpha|\alpha\rangle$. Respectively,
\begin{align}
    \llangle \underline{\alpha}| {\bf c}'_{0,n} &=\alpha_{n} \llangle \underline{\alpha} |, \ & \llangle \underline{\alpha}|{\bf c}'_{1,n} =\alpha^{*}_{n}\llangle \underline{\alpha}|,\label{eq_c1}\\
{\bf c}_{0,n}|\underline{\eta}_{a}\rrangle&=\eta^{*}_{n}|\underline{\eta}_{a}\rrangle, \ & {\bf c}_{1,n}|\underline{\eta}_{a}\rrangle=-\eta_{n}|\underline{\eta}_{a}\rrangle.\label{eq_c}
\end{align}
To make the resemblance more telling, they can be written into lookalike displaced operators acting on the superoperator vacuum for each mode

\begin{align}
\llangle \alpha_{n}|&=\llangle 0|e^{\alpha_{n}{\bf c}_{0,n}+\alpha^{*}_{n}{\bf c}_{1,n}},\\
|\eta_{a,n}\rrangle&=e^{-\eta_{n}{\bf c}'_{1,n}+\eta^{*}_{n}{\bf c}'_{0,n}}|\mathbbm{1}\rrangle.
\end{align}
The superoperator coherent states do not form an orthogonal basis but the anti-normally ordered operators do
\begin{align}
\llangle \alpha_{n} |\beta_{n} \rrangle&=e^{-|\alpha_{n} -\beta_{n}|^{2}},\\
\llangle \eta_{a,n}|\eta'_{a,n}\rrangle &=\pi e^{-|\eta_{n}|^{2}}\delta^{(2)}(\eta_{n}-\eta'_{n}),
\end{align}
with $\delta^{(2)}(\eta)=\delta ({\rm Re}\ \eta)\delta ({\rm Im}\ \eta)$. In particular, the overlap between the two superoperators is given by
\begin{align}
\llangle \alpha_{n} |\eta_{a,n}\rrangle ={\rm Tr}[|\alpha_{n}\rangle\langle \alpha_{n}|\eta_{a,n}] =e^{\eta ^{*}_{n}\alpha_{n}} e^{-\eta_{n}\alpha^{*}_{n}}e^{-|\eta_{n}|^{2} }.
\end{align}
In other words, the left and right 
eigenvectors of the superoperators are conjugated operators via Fourier transform $\mathcal{F}[\bullet]$
\begin{align}
\llangle \eta_{a,n}|\bullet\rrangle&=\mathcal{F}[\llangle \alpha_{n}|\bullet\rrangle]=\int \frac{d^{2}\alpha_{n}}{\pi}\llangle \alpha_{n}|\bullet\rrangle e^{-\eta^{*}_{n}\alpha_{n}} e^{\eta_{n}\alpha^{*}_{n}},\\
\llangle \alpha_{n}|\bullet\rrangle&=\mathcal{F}^{-1}[\llangle \eta_{a,n}|\bullet\rrangle]=\int \frac{d^{2}\eta_{n}}{\pi}\llangle \eta_{a,n}|\bullet\rrangle e^{\eta^{*}_{n}\alpha_{n}} e^{-\eta_{n}\alpha^{*}_{n}}.\label{eq:fourier_transform}
\end{align}
The Fourier transform of the anti-normally ordered characteristic function is $1/\pi$, which is the value of the $Q$ representation for a coherent state \cite{carmichael2009book}.
\section{The $Q$ representation and its characteristic function in the superoperator coherent state basis \label{Q_representation}}
The left projection of the super coherent state $\llangle \underline{\alpha}|$ on the density matrix is nothing but the $Q$ representation $Q(\underline{\alpha})=\langle \underline{\alpha}|\rho|\underline{\alpha}\rangle/\pi^{N}=\llangle \underline{\alpha} |\rho\rrangle/\pi^{N}$, which corresponds to the probability to be in the coherent state $|\underline{\alpha}\rangle$ \cite{Husimi1940}. In projecting the vectorized density matrix on \eqref{eq:fourier_transform}, the $Q$ representation is the Fourier transform of the anti-normal ordered correlation function \cite{carmichael2009book}
\begin{align}
Q(\underline{\alpha})=\int \frac{d^{2}\underline{\eta}}{\pi^{2N}}\mathcal{X}_{a}(\underline{\eta},\underline{\eta}^{*})e^{\underline{\eta}^{*}\cdot\underline{\alpha}}e^{-\underline{\eta} \cdot\underline{\alpha}^{*}},
\end{align}
\begin{table}[htbp]
\centering
\setlength\extrarowheight{4pt}
\large
\begin{tabular}{|>{\centering\arraybackslash}p{1.8cm}
                |>{\centering\arraybackslash}p{3cm}
                |>{\centering\arraybackslash}p{3cm}|} 
\hline
$x$ & $\llangle \underline{\alpha} |x|\rho\rrangle / \pi$ & $\llangle \underline{\eta}_{a} | x | \rho \rrangle$ \\
\hline
${\bf c}_{0,n}$ & $\partial_{\alpha_{n}} Q(\underline{\alpha})$ & $\eta^{*}_{n} \mathcal{X}_{a}(\underline{\eta},\underline{\eta}^{*})$ \\
\hline
${\bf c}_{1,n}$ & $\partial_{\alpha^{*}_{n}} Q(\underline{\alpha})$ & $-\eta_{n} \mathcal{X}_{a}(\underline{\eta},\underline{\eta}^{*})$ \\
\hline
${\bf c}'_{0,n}$ & $\alpha_{n} Q(\underline{\alpha})$ & $-\partial_{\eta^{*}_{n}} \mathcal{X}_{a}(\underline{\eta},\underline{\eta}^{*})$ \\
\hline
${\bf c}'_{1,n}$ & $\alpha^{*}_{n} Q(\underline{\alpha})$ & $\partial_{\eta_{n}} \mathcal{X}_{a}(\underline{\eta},\underline{\eta}^{*})$ \\
\hline
\end{tabular}

\normalsize
\caption{The table illustrates the mapping between the third quantized operators, the $Q$ representation, and its anti-normally ordered characteristic functions. The projected superoperators act in a similar way to the real space superoperator basis used to obtain the normal form of the OU process, but in the complex plane $x\bullet \to \alpha \bullet$ and $\partial_{x}\bullet \to \partial_{\alpha}\bullet$. Further details on the derivation of the equivalence table are given in Appendix~\ref{details_projection_table}.}
\label{table1}
\end{table}
with $\mathcal{X}_{a}(\underline{\eta},\underline{\eta}^{*})={\rm Tr}(\underline{\eta}^{\dagger}_{a}\rho)={\rm tr}(e^{-\underline{\eta}^{*}\cdot \underline{a}}e^{\underline{\eta}\cdot \underline{a}^{\dagger}} \rho )$, and the coherent-state amplitude vector $\underline{\alpha}^{T}=(\alpha_{1} \cdots \alpha_{N})$. The left projection on the anti-normally ordered superoperator gives the characteristic function $\llangle \underline{\eta}_{a}|\rho\rrangle=\mathcal{X}_{a}(\underline{\eta},\underline{\eta}^{*})$. Using the properties of the Fourier transform of the derivative, one obtains 
\begin{align}
\mathcal{F}[\partial_{\alpha^{*}_{n}}\llangle \alpha_{n}|]=-\eta_{n}\llangle \eta_{a,n}|,\ & \mathcal{F}[\partial_{\alpha_{n}}\llangle \alpha_{n}|]=\eta^{*}_{n}\llangle \eta_{a,n}|,\label{FT1}\\
\mathcal{F}^{-1}[\partial_{\eta^{*}_{n}}\llangle \eta_{a,n}|]=-\alpha_{n}\llangle \alpha_{n}|,\ & \mathcal{F}^{-1}[\partial_{\eta_{n}}\llangle \eta_{a,n}|]=\alpha^{*}_{n}\llangle \alpha_{n}|.\label{FT2}
\end{align}
This leads the different left projections of the superoperators ${\bf c}_{0,n},{\bf c}_{1,n},{\bf c}'_{0,n},{\bf c}'_{1,n}$, summarized in the Table~\ref{table1}. 
As a consequence, projecting the vectorized coherent state $\llangle \underline{\alpha}|$ or the vectorized antinormally ordered operator $\llangle \underline{\eta}_{a}|$ on the vectorized density matrix $|\rho\rrangle$ leads to equivalent equations of motion for the $Q$ representation and its characteristic function. More details on the derivation of table~\ref{table1} and on the projection procedure for the master equation are given in App.~\ref{details_projection_table}.

Let us now consider a quadratic Lindblad master equation of $\rm N$ bosons similar to the one introduced in \cite{Prosen_2010,McDonald2023}. The Hamiltonian includes quadratic harmonic oscillator terms and nonlinearities. The dissipator is of Lindblad form $\mathcal{D}[L](\rho)=L\rho L^{\dagger}-\frac{1}{2}\{L^{\dagger}L,\rho\}$ with a jump operator linear in the annihilation and creation of bosonic operators, leading to coherences between the pumping and loss process. The coefficients $l_{bm}$ and $p^{*}_{bm}$ are the couplings to different baths labeled by the index $b$, so that the master equation reads
\begin{widetext}
\begin{align}
\partial_{t}\rho &=-i\sum_{n,m}\mathbbm{H}_{nm}[a^{\dagger}_{n}a_{m},\rho]-\frac{i}{2}\sum_{n,m}[\mathbbm{K}_{nm}a^{\dagger}_{n}a^{\dagger}_{m}+\text{h.c},\rho]+\sum_{b}\mathcal{D}\left[\sum_{m}(l_{bm}a_{m}+p^{*}_{bm}a^{\dagger}_{m})\right](\rho) \nonumber\\
&=-i\sum_{n,m}\mathbbm{H}_{n,m}[a^{\dagger}_{n}a_{m},\rho]+\sum_{n,m}\mathbbm{L}_{nm}\left(a_{m}\rho a^{\dagger}_{n}-\frac{1}{2}\{a^{\dagger}_{n}a_{m},\rho\}\right)+\sum_{n,m}\mathbbm{P}_{nm}\left(a^{\dagger}_{n}\rho a_{m}-\frac{1}{2}\{a_{m}a^{\dagger}_{n},\rho\}\right)\nonumber\\
&-\frac{i}{2}\sum_{n,m}[\mathbbm{K}_{n,m}a^{\dagger}_{n} a^{\dagger}_{m} + \text{h.c},\rho]+\sum_{n,m}\mathbbm{C}_{n m}\left(a^{\dagger}_{m}\rho a^{\dagger}_{n}-\frac{1}{2}\{a^{\dagger}_{n}a^{\dagger}_{m},\rho\}\right)+\sum_{n,m}\mathbbm{C}^{*}_{n,m}\left(a_{n}\rho a_{m}-\frac{1}{2}\{a_{m}a_{n},\rho \}\right)=\mathcal{L}\rho.
\end{align}
\end{widetext}
The Hermiticity of the Hamiltonian implies that the matrix $\mathbbm{H}$ is Hermitian $\mathbbm{H}^{\dagger}=\mathbbm{H}$ and the nonlinearity matrix $\mathbbm{K}$ is symmetric $\mathbbm{K}=\mathbbm{K}^{T}$. The quadratic master equation is rewritten in terms of the Hermitian pumping $\mathbbm{P}_{nm}=\sum_{b}p^{*}_{bn}p_{bm}$ and loss matrices $\mathbbm{L}_{nm}=\sum_{b}l^{*}_{bn}l_{bm}$. The coherences between the loss and pumping process are embedded into the coherence matrix $\mathbbm{C}_{nm}=\sum_{b}l^{*}_{bn}p^{*}_{bm}$.
To write the master equation as a quadratic form, we introduce the superoperator vectors $\underline{\bf c}^{T}_{0}=({\bf c}_{0,1}\ {\bf c}_{0,2}\ \cdots\ {\bf c}_{0,N})$, $\underline{\bf c}^{T}_{1}=({\bf c}_{1,1}\ {\bf c}_{1,2}\ \cdots\ {\bf c}_{1,N})$ and $\underline{\bf c}'^{T}_{0}=({\bf c}'_{0,1}\ {\bf c}'_{0,2}\ \cdots\ {\bf c}'_{0,N})$, $\underline{\bf c}'^{T}_{1}=({\bf c}'_{1,1}\ {\bf c}'_{1,2}\ \cdots\ {\bf c}'_{1,N})$. The effective Hamiltonian takes the form of a non-hermitian Hamiltonian with gain and loss terms. After symmetrization, the Lindbladian is recast in a quadratic form of the vector $\underline{z}^{T}=(\underline{\bf c}'_{0} \ \underline{\bf c}'_{1} \ \underline{\bf c}_{0} \ \underline{\bf c}_{1})$. Finally, the master equation can be written
\begin{align}
\partial_{t}|\rho\rrangle&=\underline{\mathcal{L}}|\rho\rrangle,\\
\underline{\mathcal{L}}&=\frac{1}{2}\underline{z}^{T}S_{q}\underline{z}+\frac{1}{2}{\rm Tr}(\mathbbm{L}-\mathbbm{P}),\label{quantum_linblad}\\
S_{q}&=\begin{pmatrix}
0 & \rvline & \beta^{T}_{q}\\
\hline
\beta_{q} & \rvline & 2D_{q}
\end{pmatrix},\\
\beta_{q}&=\begin{pmatrix}
\mathbbm{H}_{\rm +} & \mathbbm{B}_{12}\\
\mathbbm{B}^{*}_{12} & \mathbbm{H}^{T}_{-}
\end{pmatrix},\quad D_{q}=\frac{1}{2}\begin{pmatrix}
\mathbbm{D}_{11} & \mathbbm{L}\\
\mathbbm{L}^{T} & \mathbbm{D}^{*}_{11}
\end{pmatrix}=D^{T}_{q},\\
\mathbbm{H}_{\pm}&=\pm i \mathbbm{H}+\frac{1}{2}(\mathbbm{L}-\mathbbm{P}),\\
\mathbbm{B}_{12}&=i\mathbbm{K}+\frac{1}{2}(\mathbbm{C}-\mathbbm{C}^{T}),\\
\mathbbm{D}_{11}&=i\mathbbm{K}-\frac{1}{2}(\mathbbm{C}+\mathbbm{C}^{T})=\mathbbm{D}^{T}_{11}.
\end{align}
The structure of the symmetric matrix representing the quadratic $N$-modes Lindbladian is similar to the structure of the OU process. We denote the quantum drift matrix $\beta_{q}$; its diagonal terms contain the Hamiltonian, pumping and loss matrix. Its off-diagonal terms contain the nonlinearity matrix $\mathbbm{K}$ and the coherence matrix $\mathbbm{C}$. Furthermore, whenever the coherence matrix is symmetric $\mathbbm{C}=\mathbbm{C}^{T}$, coherences do not play any role in the drift matrix and the off-diagonal drift matrix becomes symmetric $\mathbbm{B}^{T}_{12}=\mathbbm{B}_{12}$. The quantum diffusion matrix denoted $D_{q}$ is symmetric as in the OU process and its diagonal terms are governed by the nonlinearity and coherences, rather than the off diagonal terms being induced by the loss matrix.

After the projection of the vectorized coherent state on the vectorized density matrix $\llangle \underline{\alpha}|\rho\rrangle$, the equation of motion for the $Q$ representation reads
\begin{align}
\partial_{t}Q(\underline{\alpha})&=(\underline{\partial}_{\alpha} \ \underline{\partial}_{{\alpha}^{*}})\Big[\beta_{q} \begin{pmatrix} \underline{\alpha}\\
\underline{\alpha}^{*}\end{pmatrix}+D_{q} \begin{pmatrix} \underline{\partial}_{\alpha} \\
\underline{\partial}_{\alpha^{*}}
\end{pmatrix}\Big] Q(\underline{\alpha}). \label{Q_rep_equation_motion}
\end{align}
This is the equation of motion of a $2N$-dimensional complex OU process. The normal form is obtained in the same way as that for the OU process. Assuming that the matrix $\beta_{q}$ is diagonalizable \cite{Prosen_2010} and writing $\beta_{q}=P_{q}\Delta_{q} P^{-1}_{q}$, with $\Delta_{q}={\rm diag}(\beta_{1,q},\cdots,\beta_{2N,q})$, we use the known decomposition \cite{Prosen_2008} to write the Liouvillian $\underline{\mathcal{L}}$ in a normal form. $S_{q}$ is having a similar form to $S_{+}$ in the previous section so that the normal-form decomposition reads
\begin{align}
\underline{\mathcal{L}}=\frac{1}{2}\sum_{i=1}^{2N}\beta_{i,q}(\zeta'_{i,q}\zeta_{i,q}+\zeta_{i,q}\zeta'_{i,q})+\frac{1}{2}{\rm Tr}\beta_{q}.
\end{align}
where $(\underline{\zeta},\underline{\zeta}')^{T}=(V_{q}\underline{z})^{T}$,  
\begin{align}
V_{q}=\begin{pmatrix}
P^{-1}_{q}& P^{-1}_{q}\Sigma_{\alpha}(\infty)\\
0 & (P_{q})^{T}
\end{pmatrix}\label{eq:passageq}.
\end{align}
and $\Sigma_{\alpha}(\infty)$ is the solution to the Lyapunov equation 
\begin{align}
\beta_{q}\Sigma_{\alpha}(\infty)+\Sigma_{\alpha}(\infty)\beta^{T}_{q}=2D_{q}. \label{lyap_quantum}
\end{align}
The new operators take the form
\begin{align}
\zeta_{i,q}&=\sum_{j=1}^{2N}P^{-1}_{q}c'_{j}+\sum_{j=1}^{2N}(P^{-1}_{q}\Sigma_{\alpha}(\infty))_{ij}c_{j},\label{crea_quantum}\\
\zeta'_{i,q}&=\sum_{j=1}^{2N}(P^{T}_{q})_{ij} c_{j},\label{anhil_quantum}
\end{align}
where we used the notation $c_{j}=(\underline{\bf c}_{0}, \underline{\bf c}_{1})_{j}$ and  $c'_{j}=(\underline{\bf c}'_{0}, \underline{\bf c}'_{1})_{j}$. The right vacuum of this basis is the non-equilibrium steady state 
\begin{align}
\underline{\mathcal{L}}|\rho_{\rm ness}\rrangle=\zeta_{i,q}|\rho_{\rm ness}\rrangle=0.\label{eq:steady_state}
\end{align}
Given the identification of the annihilation operator, one can establish the creation operator $-\zeta'_{i,q}$ from the commutation relation $[\zeta_{i,q},-\zeta'_{j,q}]=\delta_{ij}\mathbbm{1}$. The left vacuum is the identity operator
\begin{align}
\llangle \mathbbm{1}|\underline{\mathcal{L}}=\llangle \mathbbm{1}|(-\zeta'_{i,q})=0.
\end{align}
This can be seen from the trace-preserving nature of the evolution. At the end, one can write the Lindbladian in the form
\begin{align}
\underline{\mathcal{L}}&=-\sum_{i=1}^{2N}\beta_{i,q}(-\zeta'_{i,q})\zeta_{i,q}.\label{lin_quantum}
\end{align}
One can verify that inserting Eq.~\eqref{crea_quantum} and \eqref{anhil_quantum} in Eq.~\eqref{lin_quantum} recovers the Lindbladian \eqref{quantum_linblad}. From this, one can build the eigenvector Fock basis of the Liouvillian
\begin{align}
|r_{\underline{\mu}}\rrangle &= \prod_{i=1}^{2N}\frac{(-\zeta'_{i,q})^{\mu_{i}}}{\sqrt{\mu_{i}!}}|\rho_{\rm ss}\rrangle, \
\llangle l_{\underline{\mu}}|= \llangle \mathbbm{1}|\prod_{i=1}^{2N}\frac{(\zeta_{i,q})^{\mu_{i}}}{\sqrt{\mu_{i}!}},
\end{align}
where $\mu_{i}$ represents the occupation number in the Fock basis for each mode. This leads to the spectral decompostion of the Liouvillian
\begin{align}
\underline{\mathcal{L}}|r_{\underline{\mu}}\rrangle&=E_{\underline{\mu}}|r_{\underline{\mu}}\rrangle,\ \llangle l_{\underline{\mu}}|\underline{\mathcal{L}}=E_{\underline{\mu}}\llangle l_{\underline{\mu}}|,\ E_{\underline{\mu}}=-\sum_{i=1}^{2N}\mu_{i}\beta_{i,q}.
\end{align}
The autocorrelation covariance matrix of the process divides in blocks
\begin{align}
\Sigma_{\alpha}(t)=\begin{pmatrix}\langle \underline{\alpha}(t),\underline{\alpha}^{T}(t)\rangle & \langle \underline{\alpha} (t),\underline{\alpha}^{*T}(t)\rangle\\
\langle \underline{\alpha}^{*}(t),\underline{\alpha}^{T} (t)\rangle & \langle \underline{\alpha}^{*} (t),\underline{\alpha}^{*T}(t)\rangle
\end{pmatrix}
\end{align}
with the shortcut notation for the covariance scalar product
\begin{align}
\langle \underline{\alpha}(t),\underline{\alpha}^{T}(t)\rangle=\langle [\underline{\alpha}(t)-\langle \underline{\alpha}(t)\rangle][\underline{\alpha}(t)-\langle \underline{\alpha}(t)\rangle]^{T}\rangle.
\end{align}
The nonequilibrium steady-state $Q$ representation can be obtained in solving the Lyapunov equation \eqref{eq:steady_state}. Assuming that the covariance matrix is invertible and positive definite, the associated $Q$ representation is of the form
\begin{align}
Q_{\rm ness}(\underline{\alpha})=\frac{\exp\left[-\frac{1}{2}(\underline{\alpha},\ \underline{\alpha}^{*})^{T}\Sigma^{-1}_{\alpha}(\infty)(\underline{\alpha},\ \underline{\alpha}^{*})\right]}{\pi^{N} \sqrt{{\rm det}(\Sigma_{\infty})}}.\label{eq:Q_rep_ness}
\end{align}
\section{Q representation of the eigenvectors \label{eigenvectors_Q_representation}}
One can obtain the $Q$ representation of the eigenvectors by projection on the superoperator coherent state. Their expression is similar to the real OU process, and decomposes in a sum of products of Hermite polynomials. Assuming the whitening decomposition of the inverse covariant matrix $\Sigma^{-1}_{\infty}=W^{T}_{q}W_{q}$, let us consider the change of variable
\begin{align}
\uuline{\alpha}'&=W_{q}\uuline{{\bf \alpha}},\\
\uuline{\partial}_{\alpha}&=W^{T}_{q}\uuline{\partial}_{\alpha'},
\end{align}
where the double underline is a shortcut notation for the vector composed of the coherent amplitude and its conjugate $\uuline{\alpha}=(\underline{\alpha}, \underline{\alpha}^{*})$. One can find the right eigenstate of the multidimensional OU process as further described in Appendix~\ref{app:eigenstates_Q} 
\small
\begin{align}
\llangle \alpha|r_{\underline{\mu}}\rrangle&=Q_{\rm ness}(\underline{\alpha})\sum_{\substack{k^{1}_{1}+k^{1}_{2}+\cdots+k^{1}_{2N}=\mu_{1},\\
\vdots\\
k^{2N}_{1}+k^{2N}_{2}+\cdots+k^{2N}_{2N}=\mu_{2N}}}\Big[\prod_{i=1}^{2N}\frac{1}{\sqrt{\mu_{i}!}}\binom{\mu_{i}}{k^{i}_{1},k^{i}_{2},\cdots,k^{i}_{2N}}\Big]\nonumber\\
&\times\prod_{t=1}^{2N}\Big\{\prod_{i=1}^{2N}[(W_{q}P_{q})^{T}]^{k^{i}_{t}}_{it}\Big\}H_{\sum_{i=1}^{2N}k^{i}_{t}}(\alpha'_{t}).\label{eq:eigenstate}
\end{align}
\normalsize
In particular, the NESS is recovered for  $\underline{\mu}=\underline{0}$. The left eigenvector is given by 
\begin{align}
\llangle l_{\mu}|\alpha\rrangle&=\sum_{\substack{k^{1}_{1}+k^{1}_{2}+\cdots+k^{1}_{2N}=\mu_{1},\\
\vdots\\
k^{2N}_{1}+k^{2N}_{2}+\cdots+k^{2N}_{2N}=\mu_{2N}}}\Big[\prod_{i=1}^{2N}\frac{1}{\sqrt{\mu_{i}!}}\binom{\mu_{i}}{k^{i}_{1},k^{i}_{2},\cdots,k^{i}_{2N}}\Big]\nonumber\\
&\times\prod_{t=1}^{2N}\Big\{\prod_{i=1}^{2N}(P^{-1}_{q}W^{-1}_{q})^{k^{i}_{t}}_{it}\Big\}H_{\sum_{i=1}^{2N}k^{i}_{t}}(\alpha'_{t}).\label{left_eigenstate}
\end{align}
\section{Equation of motion of the covariance matrix \label{evolution_covariance}}
The equation of motion for the symmetric covariance matrix $\Sigma_{\alpha}=\Sigma^{T}_{\alpha}$ can be obtained from the Heisenberg picture equations of motion
\begin{align}
\partial_{t}\Sigma_{\alpha}(t)&=-\beta_{q}\Sigma_{\alpha}(t)-\Sigma_{\alpha}(t)\beta^{T}_{q}+2D_{q}.
\end{align}
This equation can be solved following the vectorization procedure
\begin{align}
\partial_{t}|\Sigma_{\alpha}(t)\rrangle=-\left(\beta_{q}\otimes \mathbbm{1}_{2n}+\mathbbm{1}_{2n}\otimes \beta_{q}\right)|\Sigma_{\alpha}(t)\rrangle+|2D_{q}\rrangle,
\end{align}
with the solution 
\begin{align}
\Sigma_{\alpha}(t)=e^{-\beta_{q}t}\left[\Sigma_{\alpha}(0)-\Sigma_{\alpha}(\infty)\right]e^{-\beta^{T}_{q}t}+\Sigma_{\alpha}(\infty),
\end{align}
where the stationary covariance matrix is the solution to the Lyapunov equation \eqref{lyap_quantum}. In the vectorized form, one can write the Lyapunov equation $|2D_{q}\rrangle=(\beta_{q}\otimes \mathbbm{1}_{2n}+\mathbbm{1}_{2n}\otimes\beta_{q})|\Sigma_{\alpha}(\infty)\rrangle=(\beta_{q}\oplus \beta_{q})|\Sigma_{\alpha}(\infty)\rrangle$, so that the formal solution of the Lyapunov equation is given by $|\Sigma_{\alpha}(\infty)\rrangle=(\beta_{q}\oplus \beta_{q})^{-1}|2D_{q}\rrangle$.
\section{Evolution of the $Q$ representation \label{evolution_Q_representation}}
The evolution of the $Q$ representation is given by
\begin{align}
Q(\underline{\alpha},t)&=\llangle \alpha|e^{\underline{\mathcal{L}}t}|\rho(0)\rrangle\nonumber\\
&=\sum_{\underline{\mu}}e^{E_{\underline{\mu}}t}\llangle \underline{\alpha}|r_{\underline{\mu}}\rrangle\llangle l_{\underline{\mu}}|\rho(0)\rrangle.
\end{align}
It is not trivial to compute this expression in a closed form. Let us consider an initial coherent state $|\rho(0)\rrangle=|\underline{\alpha}_{0}\rrangle$. For simplification, let us consider the $2$D case, for which one can express the result as a sum over the generalized Hermite polynomials. Let us introduce the generalized polynomial obtained from \eqref{left_eigenstate}
\begin{align}
\mathcal{H}_{m,n}(U;x,y)&=\sum_{k=0}^{m}\sum_{j=0}^{n}\binom{m}{k}\binom{n}{j}U^{k}_{11}U^{m-k}_{12}U^{j}_{21}U^{n-j}_{22}\nonumber\\
&\times H_{k+j}(x)H_{m+n-(k+j)}(y),
\end{align}
where $U$ is a $2\times 2$ matrix and $U_{ij}$ its elements with $i$ the lines and $j$ the columns. As a consequence, one can write the evolution of the $Q$ representation for a $1$D master equation
\begin{align}
&Q(\alpha,\alpha^{*},t)=Q_{\rm ness}(\underline{\alpha})\sum_{\mu_{1},\mu_{2}=0}^{\infty}\frac{e^{-(\beta_{1}\mu_{1}+\beta_{2}\mu_{2})t}}{\mu_{1}!\mu_{2}!}\\
&\times\mathcal{H}_{\mu_{1},\mu_{2}}(-(W_{q}P_{q})^{T};\alpha',\alpha'^{*})\nonumber\mathcal{H}_{\mu_{1},\mu_{2}}((P^{-1}_{q}W^{-1}_{q});\alpha'_{0},\alpha'^{*}_{0}).
\end{align}
To obtain a closed form of this expression, one would need a formula similar to Mehler's formula used on the singular Hermite polynomials. To our knowledge, such a formula is not known. Also, we leave the general case for further work. For simplicity, we consider the simplest case $P_{q}W_{q}=\mathbbm{1}$. This simplifies the polynomial in the product of Hermite polynomials
\begin{align}
\mathcal{H}_{m,n}(x,y)&=H_{m}(x)H_{n}(y).
\end{align}
In this case, Mehler's formula is directly applicable, and the $Q$ representation greatly simplifies
\begin{align}
Q(\alpha,\alpha^{*},t)&=Q_{\rm ness}(\underline{\alpha})\sum_{\mu_{1}=0}^{\infty}\frac{e^{-\beta_{1}\mu_{1}t}}{\mu_{1}!}H_{\mu_{1}}(\alpha')H_{\mu_{1}}(\alpha'_{0})\nonumber\\
&\times\sum_{\mu_{2}=0}^{\infty}\frac{e^{-\beta_{2}\mu_{2}t}}{\mu_{2}!}H_{\mu_{2}}(\alpha'^{*})H_{\mu_{2}}(\alpha'^{*}_{0})\nonumber\\
&=Q_{\rm ness}(\underline{\alpha})E(e^{-\beta_{1}t},\alpha',\alpha'_{0})E(e^{-\beta_{2}t},\alpha'^{*},\alpha'^{*}_{0}),
\end{align}
with $E(\rho,x,y)=1/(1-\rho^{2})\exp[-[\rho^{2}(x^{2}+y^{2})-2x y\rho]/(2(1-\rho^{2}))]$. The Fock state superoperator basis leads to a cumbersome expression for the propagator; as an alternative, it may be possible to follow a path-integral approach; in inserting the resolution of the identity for the characteristic function, which would play a similar role to the path-integral coherent state approach of \cite{McDonald2023}.  
\section{Classical-Quantum correspondence in the Prosen-Seligman basis \label{prosen_seligman}}
We have demonstrated that the introduction of a new third quantized basis is useful for recasting the equation of motion of the $Q$ representation in the form of an OU process. We now turn to discuss the quantum-classical correspondence in the third quantized basis introduced by Prosen and Seligman \cite{Prosen_2010}. We demonstrate that similarly to the basis we introduce, their basis is effective for determining the equation of evolution of the $P$ representation and its associated characteristic function. For the $P$ representation, the equations of motion also take the form of an OU process. To recast the Lindbladian in a normal form, the authors introduced superoperators \cite{Prosen_2010}
\begin{align}
{\bf a}_{0,n}|\bullet\rrangle= |a_{n} \bullet\rrangle, \ & {\bf a}_{1,n}|\bullet\rrangle=|\bullet a^{\dagger}_{n}\rrangle,\\
{\bf a}'_{0,n}|\bullet\rrangle=|[a^{\dagger}_{n},\bullet]\rrangle, \ & {\bf a}'_{1,n}|\bullet\rrangle=|[a_{n},\bullet]\rrangle,
\end{align}
that verify the commutation relations 
\begin{align}
[{\bf a}_{0,n},{\bf a}'_{0,m}]=[{\bf a}_{1,n},-{\bf a}'_{1,m}]=\delta_{n,m}\mathbbm{1}\bullet,
\end{align}
and $[{\bf a}_{0,n},{\bf a}_{1,m}]=[{\bf a}'_{0,n},{\bf a}'_{1,m}]=0$. Also, for convenience, in our work we modified the sign convention for ${\bf a}'_{1,n}$ with respect to \cite{Prosen_2010} so that $({\bf a}'_{0,n})^{\dagger}={\bf a}'_{1,n}$. These ladder operators allow for the construction of a bi-orthogonal eigenbasis
\begin{align}
|r_{\mu,\nu}\rrangle&=\bigotimes_{n}\frac{({\bf a}'_{0,n})^{\mu}(-{\bf a}'_{1,n})^{\nu}}{\sqrt{\mu!}\sqrt{\nu!}}|\rho_{0}\rrangle,\\
\llangle l_{\mu,\nu}|&=\llangle \mathbbm{1}|\bigotimes_{n}\frac{({\bf a}_{0,n})^{\mu}({\bf a}_{1,n})^{\nu}}{\sqrt{\mu!}\sqrt{\nu!}},
\end{align}
and $|\rho_{0}\rrangle=||0\rangle \langle 0|\rrangle$ the vacuum state, such that $a_{n}|0\rangle=0$. Beyond their nature of ladder operators, ${\bf a}_{0,n}, {\bf a}'_{0,n}, {\bf a}_{1,n}$ and ${\bf a}'_{1,n}$ have their own eigenvectors in the Liouville space. Indeed, one can see that 
\begin{align}
    {\bf a}_{0,n}|\underline{\alpha} \rrangle =\alpha_{n} |\underline{\alpha} \rrangle, \ & {\bf a}_{1,n}|\underline{\alpha} \rrangle =\alpha^{*}_{n}|\underline{\alpha} \rrangle.
\end{align}
Furthermore, let us denote the normal ordered operator $\eta_{o,n}=e^{-\eta_{n} a^{\dagger}_{n}}e^{\eta^{*}_{n}a_{n}}$ and its tensorial product $\eta_{o}=\bigotimes_{n}\eta_{o,n}$. They both are eigenvectors of the operators ${\bf a}'_{0,n}$ and ${\bf a}'_{1,n}$
\begin{align}
{\bf a}'_{0,n}|\underline{\eta}_{o}\rrangle=-\eta^{*}_{n}|\underline{\eta}_{o}\rrangle, \ & {\bf a}'_{1,n}|\underline{\eta}_{o}\rrangle=-\eta_{n}|\underline{\eta}_{o}\rrangle, \label{eq:prosen_characteristic}
\end{align}
so that the eigenvectors of $({\bf a}_{0,n},{\bf a}_{1,n})$ and $({\bf a}'_{0,n},{\bf a}'_{1,n})$ have for overlap the characteristic function of the normally ordered characteristic function
\begin{align}
\llangle \alpha_{n} |\eta_{o,n}\rrangle=e^{-\eta_{n}\alpha^{*}_{n}}e^{\eta ^{*}_{n}\alpha_{n}}.
\end{align}
The $P$ representation is the Fourier transform of the antinormal ordered correlation function \cite{carmichael2009book}
\begin{align}
P(\underline{\alpha})&=\int \frac{d^{2}\underline{\eta}}{\pi^{2N}}\mathcal{X}_{o}(\underline{\eta},\underline{\eta}^{*})e^{\underline{\eta}^{*}\cdot\underline{\alpha}}e^{-\underline{\eta} \cdot\underline{\alpha}^{*}}, \label{def_p_rep}\\
\mathcal{X}_{o}(\underline{\eta},\underline{\eta}^{*})&=\int d^{2}\underline{\alpha}P(\underline{\alpha})e^{-\underline{\eta}^{*}\cdot\underline{\alpha}}e^{\underline{\eta}\cdot\underline{\alpha}^{*}},
\end{align}
with $\mathcal{X}_{o}(\underline{\eta},\underline{\eta}^{*})={\rm Tr}(\underline{\eta}^{\dagger}_{o}\rho)={\rm tr}(e^{\underline{\eta}\cdot \underline{a}^{\dagger}} e^{-\underline{\eta}^{*}\cdot \underline{a}}\rho )$. The left-projection on the normally ordered superoperator gives the characteristic function $\llangle \underline{\eta}_{o}|\rho\rrangle=\mathcal{X}_{o}(\underline{\eta},\underline{\eta}^{*})$. The Fourier transform of the normally ordered characteristic function for the coherent state gives the Dirac $\delta$-distribution, which is known as the $P$ representation for a coherent state \cite{carmichael2009book}.  
\begin{table}[H]
\centering
\setlength\extrarowheight{4pt}
\large

\begin{tabular}{|>{\centering\arraybackslash}p{1.8cm}
                |>{\centering\arraybackslash}p{3cm}
                |>{\centering\arraybackslash}p{3cm}|} 

\hline
$x$ 
& $\int d^{2}\underline{\alpha}\; x|\underline{\alpha}\rrangle P(\underline{\alpha})$ 
& $\llangle \underline{\eta}_{o}| x|\rho\rrangle$ \\

\hline
${\bf a}_{0,n}$ 
& $\int d^{2}\underline{\alpha}\; \alpha_{n} P(\underline{\alpha})$ 
& $-\partial_{\eta^{*}_{n}}\mathcal{X}_{o}(\underline{\eta},\underline{\eta}^{*})$ \\

\hline
${\bf a}_{1,n}$ 
& $\int d^{2}\underline{\alpha}\; \alpha^{*}_{n} P(\underline{\alpha})$ 
& $\partial_{\eta_{n}}\mathcal{X}_{o}(\underline{\eta},\underline{\eta}^{*})$ \\

\hline
${\bf a}'_{0,n}$ 
& $\int d^{2}\underline{\alpha}\; -\partial_{\alpha_{n}} P(\underline{\alpha})$ 
& $-\eta^{*}_{n}\mathcal{X}_{o}(\underline{\eta},\underline{\eta}^{*})$ \\

\hline
${\bf a}'_{1,n}$ 
& $\int d^{2}\underline{\alpha}\; \partial_{\alpha^{*}_{n}}P(\underline{\alpha})$ 
& $-\eta_{n}\mathcal{X}_{o}(\underline{\eta},\underline{\eta}^{*})$ \\

\hline
\end{tabular}
\caption{ The table illustrates the mapping between the third-quantized operators, the $P$ representation, and their associated characteristic functions. It is interesting to note that the Fourier transformations from the $P$ representation to its characteristic function of this table leads to the same result than the transformation from the $Q$ representation to its characteristic function in Table \ref{table1}; however, they result from the projection of different operators. The table for the $P$ representation can be computed by direct projection and Fourier transformation from Eq. \eqref{def_p_rep}. Further details on the derivation of the equivalence table are given in Appendix~\ref{P_rep_table}\label{table}}
\end{table}
The decomposition of the density matrix in the coherent state basis allows us to determine the equation of evolution for the $P$ representation \cite{Glauber1963} by a right projection on the superoperator coherent state 
\begin{align}
|\rho_{t}\rrangle=\int d^{2}\alpha \ P(\alpha,t)|\alpha \rrangle.
\end{align}
The equation of evolution of the characteristic function is obtained by direct projection of the normal ordered operator $\llangle \eta_{o}|$. The equivalence between the different representations is given in the Table \ref{table}. 
Note that the mapping between the third-quantized operators acting on the density matrix and the $P$-representation can also be obtained from the mapping between operators and the $P$ representation given in Ref.~\cite{Walls2008quantum}.

\section{Squeezed master equation in the three third-quantized bases and their equivalent quasi-probability distributions \label{example}}
As an example, consider the example of a single harmonic oscillator with coherences \cite{Kim1989,Gardiner1985,Agarwal1990,Kim1993,Walls2008quantum}
\begin{align}
\partial_{t}\rho&=-i[\omega a^{\dagger}a,\rho]+\Gamma_{\downarrow}\left(a\rho a^{\dagger}-\frac{1}{2}\{a^{\dagger}a,\rho\}\right)\nonumber\\
&+\Gamma_{\uparrow}\left(a^{\dagger}\rho a-\frac{1}{2}\{a a^{\dagger},\rho\}\right)+c\left(a^{\dagger}\rho a^{\dagger}-\frac{1}{2}\{a^{\dagger 2},\rho\}\right)\nonumber\\
&+c^{*}\left(a\rho a-\frac{1}{2}\{a^{2},\rho\}\right). \label{eq:ho_master equation}
\end{align}
In general, the equations of motion for the mean amplitude, mean quantum number are not easily obtained for this master equation, as they are for the damped harmonic oscillator. For this reason, it is particularly interesting to derive the equations of motion of the quasi-probability distributions. The implementation of the previous master equation \eqref{eq:ho_master equation} with coherences can be 
\begin{figure}[H]
    \centering
    \includegraphics[width=\linewidth]{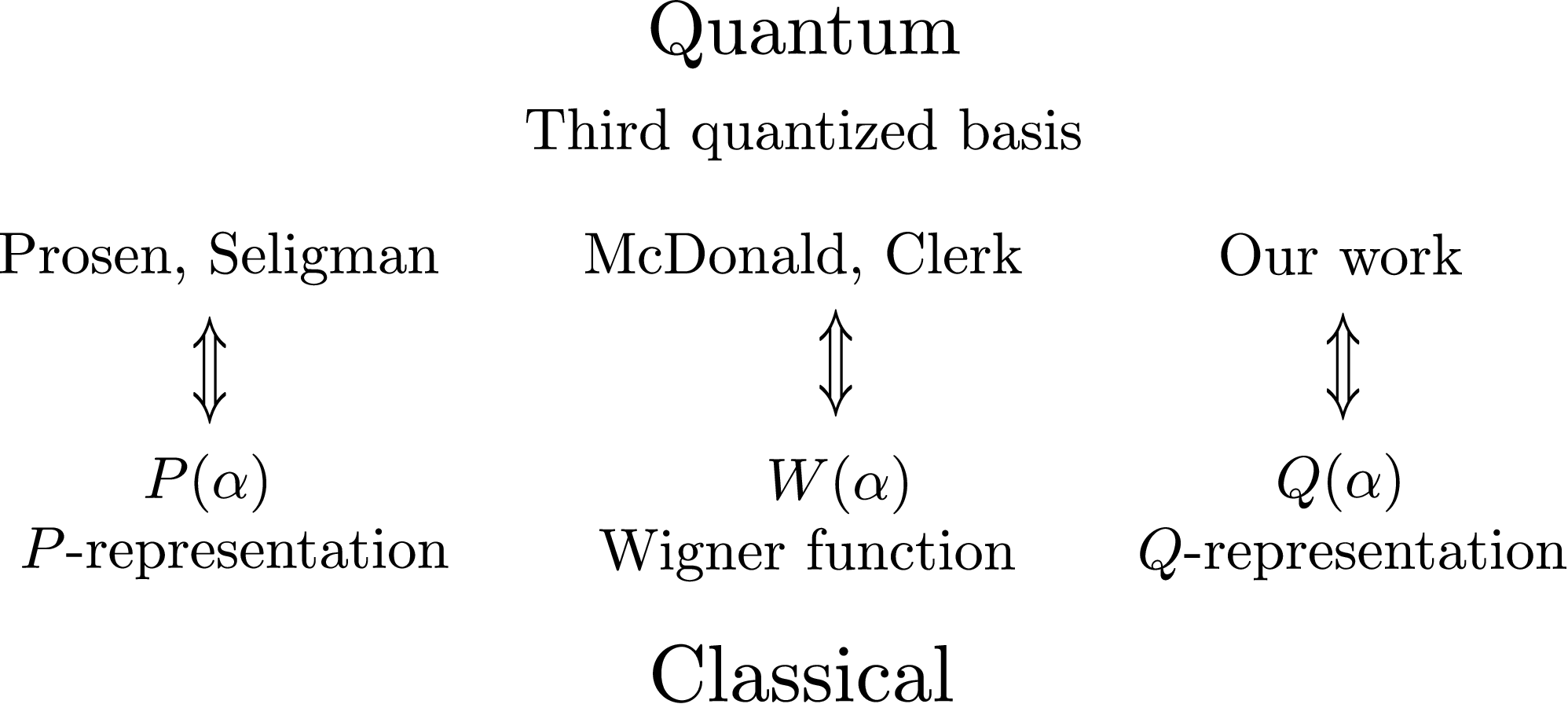}
    \caption{Different choices of third quantization basis are possible. Introducing a superoperator coherent-state basis allows us to project the third-quantized master equation into its quasiprobability distribution. We introduce a third-quantization basis to obtain the equation of motion for the $Q$ representation. Alternatively,  using the basis developed by Prosen and Seligman \cite{Prosen_2010} allows us to determine the $P$ representation equation of motion, while the basis by McDonald and Clerk allows us to determine the Wigner function representation \cite{McDonald2023}. Projecting the dynamical equation onto the quasiprobability distribution provides a clear physical interpretation in terms of an Ornstein-Uhlenbeck process.}
    \label{fig:equivalence_third_quantized}
\end{figure}
realized using a squeezed thermal bath, with coefficients given by 
\begin{align}
\Gamma_{\downarrow}&=\gamma_{\downarrow}\cosh^{2}r+\gamma_{\uparrow}\sinh^{2}r,\\
\Gamma_{\uparrow}&=\gamma_{\downarrow}\sinh^{2}r+\gamma_{\uparrow}\cosh^{2}r,\\
c&=e^{i\theta}(\gamma_{\downarrow}+\gamma_{\uparrow})\sinh(2r)/2.
\end{align}
where $r$ is the squeezing amplitude and $\theta$ the squeezing phase. For $\theta=0$ and $r=0$, there are no coherences. In this case, the master equation reduces to the simple thermal bath $\Gamma_{\downarrow}=2\kappa(\bar{n}+1)$ and $\Gamma_{\uparrow}=2\kappa\bar{n}$ and $c=0$, with $\kappa$ the coupling constant and $\bar{n}$ the mean occupation number. We would like to express the master equation Eq.~\eqref{eq:ho_master equation} in the three different third-quantized bases, and project them to their respective quasi-probability distributions, as illustrated in Fig.~\ref{fig:equivalence_third_quantized}.

For completeness, we recall the equivalence between the third-quantized basis and the Wigner function that was derived in \cite{McDonald2023}. The authors introduced the basis
\begin{align}
    {\bf a}_{\rm cl}|\rho\rrangle&=\frac{1}{\sqrt{2}}|\{a,\rho\}\rrangle,& {\bf a}_{\rm q}|\rho\rrangle&=\frac{1}{\sqrt{2}}|[a,\rho]\rrangle,\\
    {\bf a}^{\dagger}_{\rm cl}|\rho\rrangle&=\frac{1}{\sqrt{2}}|\{a^{\dagger},\rho\}\rrangle,& {\bf a}^{\dagger}_{\rm q}|\rho\rrangle&=\frac{1}{\sqrt{2}}|[a^{\dagger},\rho]\rrangle.
\end{align}
The superoperators verify the commutation relations
\begin{align}
[{\bf a}_{\rm cl},{\bf a}^{\dagger}_{\rm q}]&=[{\bf a}^{\dagger}_{\rm cl},-{\bf a}_{\rm q}]=1,\\
[{\bf a}_{\rm cl},{\bf a}^{\dagger}_{\rm cl}]&=[{\bf a}_{\rm q},{\bf a}^{\dagger}_{\rm q}]=0.
\end{align}
The eigenvectors associated to these superoperators, are equivalent to the coherent states in the vectorized representation. 
\begin{table}[H]
\centering
\setlength\extrarowheight{4pt}
\large
\begin{tabular}{|>{\centering\arraybackslash}p{1.8cm}
                |>{\centering\arraybackslash}p{3cm}
                |>{\centering\arraybackslash}p{3cm}|} 

\hline
$x$ & $\llangle \hat{\alpha}_{\rm cl}|x|\rho\rrangle$ & $\llangle \hat{\eta}_{\rm q}|x|\rho\rrangle$ \\
\hline
$\hat{\bf a}_{\rm cl}$ & $\alpha_{n} W(\alpha)$ & $\partial_{\eta^{*}_{n}} \mathcal{X}_{s}(\underline{\eta},\underline{\eta}^{*})$ \\
\hline
$\hat{\bf a}_{\rm q}$ & $\partial_{\alpha^{*}_{n}} W(\alpha)$ & $\eta_{n} \mathcal{X}_{s}(\underline{\eta},\underline{\eta}^{*})$ \\
\hline
$\hat{\bf a}^{\dagger}_{\rm cl}$ & $\alpha^{*}_{n} W(\alpha)$ & $-\partial_{\eta_{n}} \mathcal{X}_{s}(\underline{\eta},\underline{\eta}^{*})$ \\
\hline
$\hat{\bf a}^{\dagger}_{\rm q}$ & $-\partial_{\alpha_{n}} W(\alpha)$ & $\eta^{*}_{n} \mathcal{X}_{s}(\underline{\eta},\underline{\eta}^{*})$ \\
\hline

\end{tabular}

\normalsize
\caption{The table illustrates the mapping between the third-quantized operators, the Wigner representation, and their associated characteristic functions. These rules are used to obtain the equations of motion of the Wigner representation. Note that this table is reproduced from Ref.~\cite{McDonald2023}.}
\label{table_Wigner}
\end{table}

Consider the operators \cite{McDonald2023}
\begin{align}
\hat{\alpha}_{\rm cl}&\equiv 2\hat{D}(\sqrt{2}\alpha)e^{i\pi a^{\dagger}a},\\
\hat{\eta}_{q}&\equiv\hat{D}(\sqrt{2}\eta),
\end{align}
with the displacement operator $\hat{D}(\alpha)\equiv e^{\alpha \hat{a}^{\dagger}-\alpha^{*}\hat{a}}$. The displacement operator verifies the properties $a\hat{D}(\alpha)=\hat{D}(\alpha)(a+\alpha)$, and parity operator $ae^{i\pi a^{\dagger}a}=-e^{i\pi a^{\dagger}a}a$. These operators are eigenoperators for the annihilation and creation superoperators
\begin{align}
{\bf a}_{\rm cl}|\hat{\alpha}_{\rm cl}\rrangle=&\alpha |\hat{\alpha}_{\rm cl}\rrangle,& {\bf a}^{\dagger}_{\rm cl}|\hat{\alpha}_{\rm cl}\rrangle&=\alpha^{*}|\hat{\alpha}_{\rm cl}\rrangle,\\
{\bf a}_{\rm q}|\hat{\eta}_{\rm q}\rrangle&=\eta|\hat{\eta}_{\rm q}\rrangle,& {\bf a}^{\dagger}_{\rm q}|\hat{\eta}_{\rm q}\rrangle&=\eta^{*}|\hat{\eta}_{\rm q}\rrangle.
\end{align}
Consider the characteristic function 
\begin{align}
\mathcal{X}_{s}(\eta,\eta^{*})&\equiv\llangle \hat{\eta}_{\rm q}|\hat{\rho}\rrangle={\rm Tr}\left(\hat{D}^{\dagger}(\sqrt{2}\eta)\hat{\rho}\right),
\end{align}
 where the subscript $s$ stands for symmetric ordering \cite{carmichael2009book}. The Wigner function and the characteristic function are related through the Fourier transform
\begin{align}
W(\alpha)&=\int \frac{ 2d^{2}\eta}{\pi}e^{\eta \alpha^{*}-\eta^{*}\alpha}\mathcal{X}_{s}(\eta,\eta^{*}),\label{def_wigner}\\
\mathcal{X}_{s}(\eta,\eta^{*})&=\int \frac{d^{2}\alpha}{2\pi}e^{\eta^{*}\alpha-\eta\alpha^{*}}W(\alpha).\label{def_charact}
\end{align}

The moments of the Wigner function $W(\alpha)$ give the averages of operators written in Weyl order \cite{carmichael2009book}. After integration, the Wigner function can be seen as the expectation value of the parity operator \cite{Royer1977,Bishop1994}
\begin{align}
W(\alpha)&\equiv\llangle \alpha_{\rm cl}|\rho\rrangle=2{\rm Tr}\left(e^{i\pi a^{\dagger}a}\hat{D}^{\dagger}(\sqrt{2}\alpha)\hat{\rho}\right).
\end{align}
The action of the third-quantized superoperators on the density matrix $|\rho\rrangle$ was obtained in \cite{McDonald2023}. For convenience, we reproduce in Table~\ref{table_Wigner} the equivalence between the third-quantized operators and the Wigner function, as well as its characteristic function.
We are now ready to obtain the projected master equation for the Wigner function. The equivalence between the different representations for the squeezed master equation is given by
\begin{widetext}

\begin{align}
\partial_{t}|\rho\rrangle&=\Big[\left(i\omega+\frac{\Gamma_{\downarrow}-\Gamma_{\uparrow}}{2}\right){\bf c}_{0}{\bf c}'_{0}+\left(-i\omega+\frac{\Gamma_{\downarrow}-\Gamma_{\uparrow}}{2}\right){\bf c}_{1}{\bf c}'_{1}+\Gamma_{\downarrow}{\bf c}_{1}{\bf c}_{0}-\frac{c}{2}{\bf c}^{2}_{0}-\frac{c^{*}}{2}{\bf c}^{2}_{1}\Big]|\rho\rrangle,\\
\Updownarrow\nonumber\\
\partial_{t}Q(\alpha)&=\Big[\left(i\omega+\frac{\Gamma_{\downarrow}-\Gamma_{\uparrow}}{2}\right)\partial_{\alpha}\alpha+\left(-i\omega+\frac{\Gamma_{\downarrow}-\Gamma_{\uparrow}}{2}\right)\partial_{\alpha^{*}}\alpha^{*}+\Gamma_{\downarrow}\partial_{\alpha^{*}}\partial_{\alpha}-\frac{c}{2}\partial^{2}_{\alpha}-\frac{c^{*}}{2}\partial^{2}_{\alpha^{*}}\Big]Q(\alpha)\nonumber,\\
\Updownarrow\\
\partial_{t}|\rho\rrangle&=\Big[{\bf a}'_{0}{\bf a}_{0}\left(-i\omega-\frac{\Gamma_{\downarrow}-\Gamma_{\uparrow}}{2}\right)+{\bf a}'_{1}{\bf a}_{1}\left(-i\omega+\frac{\Gamma_{\downarrow}-\Gamma_{\uparrow}}{2}\right)-\Gamma_{\uparrow}{\bf a}'_{0}{\bf a}'_{1}-\frac{c}{2}{\bf a}'^{2}_{0}-\frac{c^{*}}{2}{\bf a}'^{2}_{1}\Big]|\rho\rrangle,\\
\Updownarrow\nonumber\\
\partial_{t}P(\alpha)&=\Big[\left(i\omega+\frac{\Gamma_{\downarrow}-\Gamma_{\uparrow}}{2}\right)\partial_{\alpha}\alpha+\left(-i\omega+\frac{\Gamma_{\downarrow}-\Gamma_{\uparrow}}{2}\right)\partial_{\alpha^{*}}\alpha^{*}+\Gamma_{\uparrow}\partial_{\alpha}\partial_{\alpha^{*}}-\frac{c}{2}\partial^{2}_{\alpha}-\frac{c^{*}}{2}\partial^{2}_{\alpha^{*}}\Big]P(\alpha),\\
\Updownarrow\nonumber\\
\partial_{t}|\rho\rrangle&=\Big[\left(-i\omega-\frac{\Gamma_{\downarrow}-\Gamma_{\uparrow}}{2}\right){\bf a}^{\dagger}_{\rm q}{\bf a}_{\rm cl}+\left(-i\omega+\frac{\Gamma_{\downarrow}-\Gamma_{\uparrow}}{2}\right){\bf a}_{\rm q}{\bf a}^{\dagger}_{\rm cl}-(\Gamma_{\uparrow}+\Gamma_{\downarrow}){\bf a}^{\dagger}_{\rm q}{\bf a}_{\rm q}-c{\bf a}^{\dagger 2}_{\rm q}-c^{*}{\bf a}^{2}_{\rm q}\Big]|\rho\rrangle,\\
\Updownarrow\nonumber\\
\partial_{t} W(\alpha)&=\Big[\left(i\omega+\frac{\Gamma_{\downarrow}-\Gamma_{\uparrow}}{2}\right)\partial_{\alpha}\alpha+\left(-i\omega+\frac{\Gamma_{\downarrow}-\Gamma_{\uparrow}}{2}\right)\partial_{\alpha^{*}}\alpha^{*}+(\Gamma_{\uparrow}+\Gamma_{\downarrow})\partial_{\alpha}\partial_{\alpha^{*}}-c\partial^{2}_{\alpha}-c^{*}\partial^{2}_{\alpha^{*}}\Big]W(\alpha).
\end{align}
\end{widetext}
To recover the results of \cite{Kim1993}, one can use the thermal bath parametrization $\Gamma_{\downarrow}=2\kappa(N+1)$, $\Gamma_{\uparrow}=2\kappa N$, $c=-2\kappa M$, as detailed in Appendix~\ref{consistency_check}. This also recovers the equation for the $P$ representation, the $Q$ representation, and the Wigner function written in \cite{Walls2008quantum}. The factor $2$ difference for the equation of motion of the Wigner function is due to the convention taken in \cite{McDonald2023} for the Wigner function. One can obtain the similar form to the $P$ and $Q$ representation for the squeezing term with the rescaling $\alpha \to \sqrt{2}\alpha$ and $\eta\to \eta/\sqrt{2}$ in the definitions of the Wigner function~\eqref{def_wigner} and characteristic function~\eqref{def_charact}, as detailed in App.~\ref{conventions}. The difference between the $P$, $Q$, and Wigner function representations only lies in the value of the diffusion coefficient. 

To close, let us illustrate the usefulness of our formalism by deriving the steady state for the $Q$ representation. The continuity equation \eqref{Q_rep_equation_motion} is directly expressed in terms of the drift and diffusion matrices. Consequently, with only the basic knowledge of the Hamiltonian, nonlinearity, loss, pumping, and coherence matrices, one can determine the drift and diffusion matrices. These matrices encapsulate the physics of the quadratic master equation in the form of an Ornstein–Uhlenbeck process. From them, the continuity equation for the $Q$ representation can be derived. By identification with previous notations, one has $\mathbbm{L}=\Gamma_{\downarrow}$ and $\mathbbm{P}=\Gamma_{\uparrow}$. The drift matrix is given by
\begin{align}
\beta_{q}=\begin{pmatrix}
i\omega+(\Gamma_{\downarrow}-\Gamma_{\uparrow})/2 &0\\
0&-i\omega+(\Gamma_{\downarrow}-\Gamma_{\uparrow})/2
\end{pmatrix}.
\end{align}
The diffusion matrix reads
\begin{align}
D_{q}=\frac{1}{2}\begin{pmatrix}
-c & \Gamma_{\downarrow}\\
\Gamma_{\downarrow} & -c^{*}
\end{pmatrix}.
\end{align}
Having established the drift and diffusion matrix, the associated continuity equation is directly obtained from the more general equation \eqref{Q_rep_equation_motion}
\begin{align}
\partial_{t}Q(\underline{\alpha})+\uline{\nabla}^{T}J(\underline{\alpha},t)&=0,\\
J(\underline{\alpha},t)&=-[\beta_{q}\uline{\alpha}+D_{q}\uline{\nabla}{\rm ln}Q(\underline{\alpha},t)]Q(\underline{\alpha},t).
\end{align}
with $\uline{\alpha}^{T}=(\alpha \ \alpha^{*})$, $\uline{\nabla}^{T}=(\partial_{\alpha} \ \partial_{\alpha^{*}})$, and $J(\underline{\alpha},t)$ is the current. The NESS of this continuity equation takes the form 
\begin{align}
Q_{\rm ness}(\underline{\alpha})=\frac{\exp\left[-\frac{1}{2}\uline{\alpha}^{T}\Sigma^{-1}_{\alpha}(\infty)\uline{\alpha}\right]}{\pi \sqrt{{\rm det}(\Sigma_{\alpha}(\infty))}},
\end{align}
where the solution to the Lyapunov equation \eqref{lyap_quantum} can be found as
\begin{align}
(\Sigma_{\alpha}(\infty))_{11}&=\frac{c}{-2i\omega+\Gamma_{\uparrow}-\Gamma_{\downarrow}},\\
(\Sigma_{\alpha}(\infty))_{12}&=\frac{\Gamma_{\downarrow}}{\Gamma_{\downarrow}-\Gamma_{\uparrow}},\\
(\Sigma_{\alpha}(\infty))_{22}&=\frac{c^{*}}{2i\omega+\Gamma_{\uparrow}-\Gamma_{\downarrow}}.
\end{align}
The obtained steady state is the same to that found in the literature \cite{Walls2008quantum, Mattos2025}, as further detailed in Appendix~\ref{consistency_check} . 

\section{ Conclusion \label{conclusion}}
In this article, we introduced a new basis for third quantization that is proper to derive the equations of motion of the $Q$-representation. Importantly, we showed that the equation of motion for the $Q$ representation of the quadractic master equation conforms to an OU process, and that the covariance matrix can be directly obtained from the OU parameters, namely the drift coefficient matrix and the diffusion matrix by solving the Lyapunov equation. Our work opens the door for several investigations. It demonstrates that quadratic Lindblad master equation can be described using an OU process, and its dynamics can be simulated with simple colored noise in the laboratory. The driven OU process is prevalent in active matter \cite{Martin2021}, and our results could be useful to transpose classical results in the quantum realm. Moreover, the quantum-classical correspondence in our work is not restricted to quadratic master equations; it would be interesting to explore its application to nonlinear master equations. Last but not least, the $Q$ representation is particularly well suited for studying the entropy of the system, as it remains positive. The von Neumann entropy in the $Q$ representation is known as the Wehrl entropy \cite{Wehrl1979}. On the basis of this, one can define the entropy production and heat dissipation rates, which are studied in the context of quantum thermodynamics. In further work, it would be interesting to apply the definitions of the entropy production rate and heat dissipation known for the OU process to the $Q$-representation \cite{Qian2001,Qian2005,Qian2006}. 

{\it Acknowledgement.---} We thank Adolfo del Campo for suggestions on the manuscript. 

{\it Note added.---}After completion of this work, the author discovered reference \cite{Leen2016}, that builds the eigenvectors of the OU process with ladder operators that are similar to the ones obtained from the normal form decomposition. 
\appendix
\section{Normal form solution \label{normal_form_OU}}
In this section, we provide further details to determine the normal form decomposition of a symmetric squared matrix. We use the solution suggested in \cite{Prosen_2010} to perform the canonical transformation on the symmetric matrix $S$ \eqref{eq:matrix_symmetric}, in assuming $\beta$ to be diagonalizable. 

Let us introduce the symplectic unit matrix $J=i\sigma_{y}\otimes \mathbbm{1}_{N}$, that explicitly reads
\begin{align}
J=\begin{pmatrix}
0 & \mathbbm{1}_{N}\\
-\mathbbm{1}_{N} & 0
\end{pmatrix}.
\end{align}
We would like to determine the normal form of $S$. To do so, it is necessary to diagonalize $JS$, because $S$ is symmetric, $JS$ belongs to the symplectic algebra $\mathfrak{sp}(N,\mathbbm{R})$ since $J(JS)+(JS)^{T}J=0$. As a consequence, there exists an invertible passage matrix $V$ so that $JS=V^{-1}DV$ and its eigenvalues come by pairs $\{\beta_{i},-\beta_{i}\}$ \cite{Kustura2019}. One can then look for the spectral decomposition of $JS$ in the form
\begin{align}
JS_{-}=V^{-1}\begin{pmatrix}
-\Delta &0\\
0 &\Delta
\end{pmatrix}V=V^{-1}D_{-}V,
\end{align}
and $\Delta={\rm diag}(\beta_{1},\cdots,\beta_{N})$. Assuming that the matrix $\beta$ is diagonalizable \cite{Prosen_2010} and writing $\beta=P\Delta P^{-1}$, it is possible to determine the form of $V$, as given in the main text \eqref{eq:passage}. Interestingly, $V$ satisfies the symplectic condition $V^{T} J V= J$ that helps to proceed to the symplectic decomposition
\begin{align}
\underline{{\bf c}}^{T}S_{-}\underline{{\bf c}}&=\underline{{\bf c}}^{T}J^{T}JS_{-}\underline{{\bf c}}\nonumber\\
&=\underline{{\bf c}}^{T}V^{T}J^{T}VV^{-1}D_{-}V\underline{{\bf c}}\nonumber\\
&=(V\underline{{\bf c}})^{T}(J^{T}D_{-})(V\underline{{\bf c}}).
\end{align}
The matrix 
\begin{align}
J^{T}D_{-}=-\begin{pmatrix}
0 &\Delta\\
\Delta & 0
\end{pmatrix},
\end{align}
is known as the normal form of $S_{-}$. One would like to determine the form of the matrix $V$, that diagonalizes the matrix $S_{-}$
\begin{align}
VJS_{-} V^{-1}=\begin{pmatrix}-\Delta & 0\\
0 & \Delta\end{pmatrix}.
\end{align}
One can assume the block triangular structure and its inverse
\begin{align}
V=\begin{pmatrix}
A & B\\
0 & C
\end{pmatrix},\quad
V^{-1}=\begin{pmatrix}
A^{-1} & -A^{-1}BC^{-1}\\
0 & C^{-1}
\end{pmatrix},
\end{align}
leading to the following system of equations
\begin{align}
-A\beta A^{-1}&=-\Delta,\\
C\beta^{T} C^{-1}&=\Delta,\\
\beta A^{-1}B+A^{-1}B\beta^{T}+2D&=0.
\end{align}
The system is solved by
\begin{align}
A&=P^{-1},\\
\beta Z+Z\beta^{T}&=2D,\\
B&=-P^{-1}Z,\\
C&=P^{T},
\end{align}
recovering the passage matrix of the main text \eqref{eq:passage}.
\section{Normal form solution for the adjoint operator}
We look for the spectral decomposition of $JS_{+}$ in the form
\begin{align}
JS_{+}=(V^{+})^{-1}\begin{pmatrix}
\Delta &0\\
0 &-\Delta
\end{pmatrix}V^{+}=(V^{+})^{-1}D_{+}V^{+},
\end{align}
and $\Delta={\rm diag}(\beta_{1},\cdots,\beta_{N})$. Assuming that the matrix $\beta$ is diagonalizable \cite{Prosen_2010} and writing $\beta=P\Delta P^{-1}$, it is possible to determine the form of $V^{+}$, as given in the main text \eqref{eq:passage_plus}. Interestingly, $V^{+}$ satisfies the symplectic condition $(V^{+})^{T} J V^{+}= J$ that helps to proceed to the symplectic decomposition
\begin{align}
\underline{{\bf c}}^{T}S_{+}\underline{{\bf c}}&=\underline{{\bf c}}^{T}J^{T}JS_{+}\underline{{\bf c}}\nonumber\\
&=\underline{{\bf c}}^{T}(V^{+})^{T}J^{T}V^{+}(V^{+})^{-1}D_{+}V^{+}\underline{{\bf c}}\nonumber\\
&=(V^{+}\underline{{\bf c}})^{T}(J^{T}D_{+})(V^{+}\underline{{\bf c}}).
\end{align}
The matrix 
\begin{align}
J^{T}D_{+}=\begin{pmatrix}
0 &\Delta\\
\Delta & 0
\end{pmatrix},
\end{align}
is known as the normal form of $S_{+}$. One would like to determine the form of the matrix $V^{+}$, that diagonalizes the matrix $S_{+}$
\begin{align}
V^{+}JS_{+} (V^{+})^{-1}=\begin{pmatrix}\Delta & 0\\
0 & -\Delta\end{pmatrix}.
\end{align}
One can assume the bloc triangular structure and its inverse
\begin{align}
V^{+}=\begin{pmatrix}
A & B\\
0 & C
\end{pmatrix},\quad
(V^{+})^{-1}=\begin{pmatrix}
A^{-1} & -A^{-1}BC^{-1}\\
0 & C^{-1}
\end{pmatrix},
\end{align}
leading to the following system of equations
\begin{align}
A\beta A^{-1}&=\Delta,\\
-C\beta^{T} C^{-1}&=-\Delta,\\
\beta A^{-1}B+A^{-1}B\beta^{T}&=2D.
\end{align}
The system is solved by
\begin{align}
A&=P^{-1},\\
\beta Z+Z\beta^{T}&=2D,\\
B&=P^{-1}Z,\\
C&= P^{T},
\end{align}
recovering the passage matrix of the main text \eqref{eq:passage_plus}.
\section{Details on the derivation of the eigenstates of the Ornstein-Uhlenbeck process \label{app:eigenstates}}
Let us consider the change of variable
\begin{align}
\underline{x}'&=W\underline{x},\\
\underline{\partial}_{x}&=W^{T}\underline{\partial}_{x'}.
\end{align}
The right eigenvector is expressed as
\begin{widetext}
\begin{align}
r_{\underline{\mu}}(\underline{x})&=\prod_{i=1}^{N}\frac{(\zeta'_{i})^{\mu_{i}}}{\sqrt{\mu_{i}!}}P_{\rm ness}(\underline{x}),\\
&=\prod_{i=1}^{N}\frac{[-\sum_{j=1}^{N}(WP)^{T}_{ij}\partial_{x'_{j}}]^{\mu_{i}}}{\sqrt{\mu_{i}!}}P_{\rm ness}(\underline{x}),\\
&=\prod_{i=1}^{N}\Big\{\frac{1}{\sqrt{\mu_{i}!}}\sum_{\substack{k^{i}_{1}+k^{i}_{2}+\cdots+k^{i}_{N}=\mu_{i}, \\ k^{i}_{1},k^{i}_{2},\cdots,k^{i}_{N}\geq 0}}\binom{\mu_{i}}{k^{i}_{1},k^{i}_{2},\cdots,k^{i}_{N}}\prod_{t=1}^{N}[-(WP)^{T}]^{k^{i}_{t}}_{it}\partial^{k^{i}_{t}}_{x'_{t}}\Big\}P_{\rm ness}(\underline{x}),\\
&=\sum_{\substack{k^{1}_{1}+k^{1}_{2}+\cdots+k^{1}_{N}=\mu_{1},\\
\vdots\\
k^{N}_{1}+k^{N}_{2}+\cdots+k^{N}_{N}=\mu_{N}}}\prod_{i=1}^{N}\Big\{\frac{1}{\sqrt{\mu_{i}!}}\binom{\mu_{i}}{k^{i}_{1},k^{i}_{2},\cdots,k^{i}_{N}}\prod_{t=1}^{N}[-(WP)^{T}]^{k^{i}_{t}}_{it}\partial^{k^{i}_{t}}_{x'_{t}}\Big\}P_{\rm ness}(\underline{x}),\\
&=\sum_{\substack{k^{1}_{1}+k^{1}_{2}+\cdots+k^{1}_{N}=\mu_{1},\\
\vdots\\
k^{N}_{1}+k^{N}_{2}+\cdots+k^{N}_{N}=\mu_{N}}}\Big[\prod_{i=1}^{N}\frac{1}{\sqrt{\mu_{i}!}}\binom{\mu_{i}}{k^{i}_{1},k^{i}_{2},\cdots,k^{i}_{N}}\Big]\prod_{t=1}^{N}\Big\{\prod_{i=1}^{N}[-(WP)^{T}]^{k^{i}_{t}}_{it}\Big\}\partial^{\sum_{i=1}^{N}k^{i}_{t}}_{x'_{t}}P_{\rm ness}(\underline{x}),\\
&=\sum_{\substack{k^{1}_{1}+k^{1}_{2}+\cdots+k^{1}_{N}=\mu_{1},\\
\vdots\\
k^{N}_{1}+k^{N}_{2}+\cdots+k^{N}_{N}=\mu_{N}}}\Big[\prod_{i=1}^{N}\frac{1}{\sqrt{\mu_{i}!}}\binom{\mu_{i}}{k^{i}_{1},k^{i}_{2},\cdots,k^{i}_{N}}\Big]\prod_{t=1}^{N}\Big\{\prod_{i=1}^{N}[-(WP)^{T}]^{k^{i}_{t}}_{it}\Big\}\partial^{\sum_{i=1}^{N}k^{i}_{t}}_{x'_{t}}\frac{\exp(-\frac{1}{2}\underline{x}'^{T}\underline{x}')}{(2\pi)^{N/2}({\rm det}\Sigma_{\infty})^{1/2}},\\
&=P_{\rm ness}(\underline{x})\sum_{\substack{k^{1}_{1}+k^{1}_{2}+\cdots+k^{1}_{N}=\mu_{1},\\
\vdots\\
k^{N}_{1}+k^{N}_{2}+\cdots+k^{N}_{N}=\mu_{N}}}\Big[\prod_{i=1}^{N}\frac{1}{\sqrt{\mu_{i}!}}\binom{\mu_{i}}{k^{i}_{1},k^{i}_{2},\cdots,k^{i}_{N}}\Big]\prod_{t=1}^{N}\Big\{\prod_{i=1}^{N}[(WP)^{T}]^{k^{i}_{t}}_{it}\Big\}H_{\sum_{i=1}^{N}k^{i}_{t}}(x'_{t}),
\end{align}
\end{widetext}
where we used the multinomial formula, and the definition of the probabilist Hermite polynomials $e^{-x^{2}_{t}/2}H_{n}(x_{t})=(-1)^{n}\partial^{n}_{x_{t}}e^{-x^{2}_{t}/2}$.
Let us now consider the case of the left eigenstate. The left eigenstate of the forward propagator is equal to the right eigenstate of the backward propagator, so that $\underline{L}^{+}l_{\underline{\mu}}(\underline{x})=E_{\underline{\mu}}l_{\underline{\mu}}(\underline{x})$. One can consider the creation operator for the backward propagator in the $\underline{x}'$ basis
\begin{align}
\underline{\zeta}'^{+}&=P^{-1}\underline{x}-P^{-1}\Sigma_{\infty}\underline{\partial}_{x}\\
&=P^{-1}W^{-1}\underline{x}'-P^{-1}W^{-1}W\Sigma_{\infty}W^{T}\underline{\partial}_{x'}\\
&=P^{-1}W^{-1}[\underline{x}'-\underline{\partial}_{x'}],
\end{align}
where we used the whitening property $W\Sigma_{\infty}W^{T}=\mathbbm{1}$. Furthermore, $\forall i\neq j$, $[x'_{i}-\partial'_{x_{i}},x'_{j}-\partial_{x'_{j}}]=0$, so that one can use the multivariate binomial formula and
\begin{widetext}
\begin{align}
l_{\mu}(\underline{x})&=\prod_{i=1}^{N}\frac{\left(\sum_{j=1}^{N}(P^{-1}W^{-1})_{ij}[x'_{j}-\partial_{x'_{j}}]\right)^{\mu_{i}}}{\sqrt{\mu_{i}!}}1\\
&=\prod_{i=1}^{N}\frac{1}{\sqrt{\mu_{i}!}}\sum_{k^{i}_{1}+\cdots+k^{i}_{N}=\mu_{i}}\binom{\mu_{i}}{k^{i}_{1},\cdots,k^{i}_{N}}\prod_{t=1}^{N}(P^{-1}W^{-1})^{k^{i}_{t}}_{it}[x'_{t}-\partial_{x'_{t}}]^{k^{i}_{t}}1\\
&=\sum_{\substack{k^{1}_{1}+k^{1}_{2}+\cdots+k^{1}_{N}=\mu_{1},\\
\vdots\\
k^{N}_{1}+k^{N}_{2}+\cdots+k^{N}_{N}=\mu_{N}}}\prod_{i=1}^{N}\Big\{\frac{1}{\sqrt{\mu_{i}!}}\binom{\mu_{i}}{k^{i}_{1},k^{i}_{2},\cdots,k^{i}_{N}}\prod_{t=1}^{N}(P^{-1}W^{-1})^{k^{i}_{t}}_{it}[x'_{t}-\partial_{x'_{t}}]^{k^{i}_{t}}\Big\}1\\
&=\sum_{\substack{k^{1}_{1}+k^{1}_{2}+\cdots+k^{1}_{N}=\mu_{1},\\
\vdots\\
k^{N}_{1}+k^{N}_{2}+\cdots+k^{N}_{N}=\mu_{N}}}\Big[\prod_{i=1}^{N}\frac{1}{\sqrt{\mu_{i}!}}\binom{\mu_{i}}{k^{i}_{1},k^{i}_{2},\cdots,k^{i}_{N}}\Big]\prod_{t=1}^{N}\Big\{\prod_{i=1}^{N}(P^{-1}W^{-1})^{k^{i}_{t}}_{it}\Big\}[x'_{t}-\partial_{x'_{t}}]^{\sum_{i=1}^{N}k^{i}_{t}}1\\
&=\sum_{\substack{k^{1}_{1}+k^{1}_{2}+\cdots+k^{1}_{N}=\mu_{1},\\
\vdots\\
k^{N}_{1}+k^{N}_{2}+\cdots+k^{N}_{N}=\mu_{N}}}\Big[\prod_{i=1}^{N}\frac{1}{\sqrt{\mu_{i}!}}\binom{\mu_{i}}{k^{i}_{1},k^{i}_{2},\cdots,k^{i}_{N}}\Big]\prod_{t=1}^{N}\Big\{\prod_{i=1}^{N}(P^{-1}W^{-1})^{k^{i}_{t}}_{it}\Big\}H_{\sum_{i=1}^{N}k^{i}_{t}}(x'_{t}),
\end{align}
\end{widetext}
where the probabilist Hermite polynomial is given by the Dynkin formula $H_{n}(x)=\left(x-\partial_{x}\right)^{n}1=(-1)^{n}e^{\frac{x^{2}}{2}}\frac{d^{n}}{dx^{n}}e^{-\frac{x^{2}}{2}}$. Importantly, one can see that the eigenfunctions that form the basis for the eigenvectors is a product of $N$ Hermite polynomials. 
\section{Derivation of the Q representation eigenstates \label{app:eigenstates_Q}}
Let us consider the change of variable
\begin{align}
\uuline{\alpha}'&=W_{q}\uuline{\alpha},\\
\uuline{\partial}_{\alpha}&=W^{T}_{q}\uuline{\partial}_{\alpha'}.
\end{align}
with $W_{q}$ the $2N\times 2N$ whitening matrix. The $Q$ representation of the eigenstate is given by
\begin{widetext}
\begin{align}
\llangle \underline{\alpha}|r_{\underline{\mu}}\rrangle&=\prod_{i=1}^{2N}\frac{(-\zeta'_{i,q})^{\mu_{i}}}{\sqrt{\mu_{i}!}}Q_{\rm ness}(\underline{\alpha})\\
&=\prod_{i=1}^{2N}\frac{[-\sum_{j=1}^{2N}(W_{q}P_{q})^{T}_{ij}\partial_{\alpha'_{j}}]^{\mu_{i}}}{\sqrt{\mu_{i}!}}Q_{\rm ness}(\underline{\alpha})\\
&=\prod_{i=1}^{2N}\frac{1}{\sqrt{\mu_{i}!}}\sum_{\substack{k^{i}_{1}+k^{i}_{2}+\cdots+k^{i}_{2N}=\mu_{i}, \\ k^{i}_{1},k^{i}_{2},\cdots,k^{i}_{2N}\geq 0}}\binom{\mu_{i}}{k^{i}_{1},k^{i}_{2},\cdots,k^{i}_{2N}}\nonumber\prod_{t=1}^{2N}[-(W_{q}P_{q})^{T}]^{k^{i}_{t}}_{it}\partial^{k^{i}_{t}}_{\alpha'_{t}}Q_{\rm ness}(\underline{\alpha})\\
&=\sum_{\substack{k^{1}_{1}+k^{1}_{2}+\cdots+k^{1}_{N}=\mu_{1},\\
\vdots\\
k^{N}_{1}+k^{N}_{2}+\cdots+k^{N}_{N}=\mu_{N}}}\prod_{i=1}^{N}\Big\{\frac{1}{\sqrt{\mu_{i}!}}\binom{\mu_{i}}{k^{i}_{1},k^{i}_{2},\cdots,k^{i}_{N}}\prod_{t=1}^{N}[-(W_{q}P_{q})^{T}]^{k^{i}_{t}}_{it}\partial^{k^{i}_{t}}_{\alpha'_{t}}\Big\}Q_{\rm ness}(\underline{\alpha})\\
&=\sum_{\substack{k^{1}_{1}+k^{1}_{2}+\cdots+k^{1}_{N}=\mu_{1},\\
\vdots\\
k^{N}_{1}+k^{N}_{2}+\cdots+k^{N}_{N}=\mu_{N}}}\Big[\prod_{i=1}^{N}\frac{1}{\sqrt{\mu_{i}!}}\binom{\mu_{i}}{k^{i}_{1},k^{i}_{2},\cdots,k^{i}_{N}}\Big]\prod_{t=1}^{N}\Big\{\prod_{i=1}^{N}[-(W_{q}P_{q})^{T}]^{k^{i}_{t}}_{it}\partial^{\sum_{i=1}^{N}k^{i}_{t}}_{\alpha'_{t}}\Big\}\frac{\exp(-\frac{1}{2}\uuline{\alpha}'^{T}\uuline{\alpha}')}{(2\pi)^{N/2}({\rm det}\Sigma_{\infty})^{1/2}}\\
&=Q_{\rm ness}(\underline{\alpha})\sum_{\substack{k^{1}_{1}+k^{1}_{2}+\cdots+k^{1}_{2N}=\mu_{1},\\
\vdots\\
k^{2N}_{1}+k^{2N}_{2}+\cdots+k^{2N}_{2N}=\mu_{2N}}}\Big[\prod_{i=1}^{2N}\frac{1}{\sqrt{\mu_{i}!}}\binom{\mu_{i}}{k^{i}_{1},k^{i}_{2},\cdots,k^{i}_{2N}}\Big]\prod_{t=1}^{2N}\Big\{\prod_{i=1}^{2N}[(W_{q}P_{q})^{T}]^{k^{i}_{t}}_{it}\Big\}H_{\sum_{i=1}^{2N}k^{i}_{t}}(\alpha'_{t}).
\end{align}
\end{widetext}
Let us now treat the case of the left eigenstates. Consider the change of variable 
\begin{align}
\tilde{c}_{k}&=W_{q}c_{k}\\
c'_{k}&=W^{T}_{q}\tilde{c}'_{k}.
\end{align}
Applying this whitening transformation, the annihilation operator reads
\begin{align}
\zeta_{i}&=\sum_{k=1}^{2N}[P^{-1}_{q}]_{ik}c_{k}+\sum_{k=1}^{2N}[P^{-1}_{q}\Sigma_{\infty}(\alpha)]_{ik}c'_{k}\\
&=\sum_{k=1}^{2N}[P^{-1}_{q}W^{-1}_{q}]_{ik}(\tilde{c}_{k}+\tilde{c}'_{k})
\end{align}
Using the multinomial formula
\begin{align}
\llangle l_{\mu}|\alpha\rrangle&=\llangle \mathbbm{1}|\prod_{i=1}^{2N}\frac{(\zeta_{i,q})^{\mu_{i}}}{\sqrt{\mu_{i}!}}|\alpha\rrangle\\
&=\prod_{i=1}^{2N}\sum_{k^{i}_{1}+\cdots+k^{i}_{2N}=\mu_{i}}\frac{1}{\sqrt{\mu_{i}!}}\binom{\mu_{i}}{k^{i}_{1},\cdots,k^{i}_{2N}}\\
&\times\prod_{t=1}^{2N}[P^{-1}_{q}W^{-1}_{q}]^{k^{i}_{t}}_{it}\llangle \mathbbm{1}|(\tilde{c}_{t}+\tilde{c}'_{t})^{k^{i}_{t}}|\alpha\rrangle.\label{eq_interme_weyl}
\end{align}
However, $\tilde{c}_{t}$ and $\tilde{c}'_{t}$ obey the Weyl algebra from their commutation relation $[\tilde{c}_{t},\tilde{c}'_{t}]=1$. One can use the binomial theorem for noncommutative operators that obey the Weyl algebra \cite{Mikhailov1983,Varvak2005}
\begin{align}
\llangle \mathbbm{1}|(\tilde{c}_{t}+\tilde{c}'_{t})^{k_{t}}|\alpha\rrangle&=\sum_{m=0}^{k_{t}}\sum_{j=0}^{{\rm min}(m,k_{t}-m)}\frac{k_{t}!}{2^{j}j!(m-j)!(k_{t}-m-j)!}\nonumber\\
&\times{\rm Tr}\left[\tilde{c}^{m-j}_{t}\tilde{c}'^{k_{t}-m-j}_{t}|\alpha\rangle\langle \alpha|\right]\\
&=\sum_{m=0}^{n}\sum_{j=0}^{{\rm min}(m,k_{t}-m)}\frac{k_{t}!}{2^{j}j!(m-j)!(k_{t}-m-j)!}\nonumber\\
&\times{\rm Tr}\left[\tilde{c}^{m-j}_{t}|\alpha\rangle\langle \alpha|\tilde{a}^{k_{t}-m-j}_{t}\right]\\
&=\sum_{m=0}^{n}\sum_{j=0}^{{\rm min}(m,k_{t}-m)}\frac{k_{t}!}{2^{j}j!(m-j)!(k_{t}-m-j)!}\nonumber\\
&{\rm Tr}\left[(-\partial_{\alpha'_{t}})^{m-j}|\alpha\rangle \langle \alpha |\tilde{a}^{k_{t}-m-j}_{t}\right]\\
&=\sum_{m=0}^{n}\sum_{j=0}^{{\rm min}(m,k_{t}-m)}\frac{k_{t}!}{2^{j}j!(m-j)!(k_{t}-m-j)!}\nonumber\\
&(-\partial_{\alpha'_{t}})^{m-j}{\rm Tr}\left[\alpha'^{k_{t}-m-j}_{t}|\alpha\rangle \langle \alpha |\right]\\
&=(\alpha'_{t}-\partial_{\alpha'_{t}})^{k_{t}}{\rm Tr}[|\alpha\rangle\langle \alpha|].
\end{align}
where we denote the eigenvalue of the coherent state $\alpha'_{t}$ as it was taken in the whitened basis $\tilde{a}_{t}|\alpha\rangle=\alpha'_{t}|\alpha\rangle$. Making use of this expression in \eqref{eq_interme_weyl}, one can pursue the calculations
\begin{widetext}
\begin{align}
\llangle l_{\mu}|\alpha\rrangle&=\prod_{i=1}^{2N}\sum_{k_{1}+\cdots+k_{2N}=\mu_{i}}\frac{1}{\sqrt{\mu_{i}!}}\binom{\mu_{i}}{k^{i}_{1},\cdots,k^{i}_{2N}}\prod_{t=1}^{2N}[P^{-1}_{q}W^{-1}_{q}]^{k^{i}_{t}}_{it}[\alpha'_{t}-\partial_{\alpha'_{t}}]^{k^{i}_{t}}1,\\
&=\sum_{\substack{k^{1}_{1}+k^{1}_{2}+\cdots+k^{1}_{2N}=\mu_{1}\\
\vdots\\
k^{2N}_{1}+k^{2N}_{2}+\cdots+k^{2N}_{2N}=\mu_{2N}}}\Big[\prod_{i=1}^{2N}\frac{1}{\sqrt{\mu_{i}!}}\binom{\mu_{i}}{k^{i}_{1},k^{i}_{2},\cdots,k^{i}_{2N}}\Big]\prod_{t=1}^{2N}\Big\{\prod_{i=1}^{2N}(P^{-1}_{q}W^{-1}_{q})^{k^{i}_{t}}_{it}\Big\}[\alpha'_{t}-\partial_{\alpha'_{t}}]^{\sum_{i=1}^{2N}k^{i}_{t}}1.
\end{align}
\end{widetext}
Finally, we use the Dynkin formula to get the final form in the main text \eqref{left_eigenstate}.
\section{Ladder operators in function of the eigenvectors of the drift matrix}
In this section we express the ladder operators in function of the right and left eigenvectors of the drift matrix. In the main text, we assumed that the drift matrix is diagonalizable with passage matrix $P$
\begin{align}
\beta&=P\Delta P^{-1}.
\end{align}
The column vectors of $P$ are the right eigenvectors of $\beta$
\begin{align}
P=\begin{pmatrix}e_{1} &\cdots & e_{N}\end{pmatrix},
\end{align}
where we denote by $e_{i}$, and $i\in \{1,\cdots,N\}$, the eigenvector of $\beta$, such that
\begin{align}
\beta e_{i}=\beta_{i}e_{i}.
\end{align}
This is seen from the equation
\begin{align}
P\beta&=P\Delta.
\end{align}
Furthermore, the row vectors of $P^{-1}$ are the left eigenvectors of $\beta$, denoted $w_{i}$, with $i\in \{1,\cdots,N\}$
\begin{align}
w^{T}_{i}\beta&=\beta_{i}w^{T}_{i},
\end{align}
such that 
\begin{align}
P^{-1}=\begin{pmatrix}w^{T}_{1}\\ \vdots \\ w^{T}_{N}\end{pmatrix}.
\end{align}
Note that $w_{i}$ is considered as a column vector so that $w^{T}_{i}$ is a row. This is seen from the equation
\begin{align}
P^{-1}\beta&=\Delta P^{-1}.
\end{align}
As a consequence, one can rewrite the creation and annihilation operators as a function of the eigenvectors of the drift matrix
\begin{align}
\zeta_{i}&=\sum_{j}P^{-1}_{ij}x_{j}+\sum_{j}(P^{-1}\Sigma_{\infty})_{ij}\partial_{x_{j}}=w_{i}\cdot \underline{x}+(\Sigma_{\infty}w_{i})\cdot \underline{\partial}_{x},\\
\zeta'_{i}&=-\sum_{j}P^{T}_{ij}\partial_{x_{j}}=-e_{i}\cdot\underline{\partial}_{x},
\end{align}
with the scalar product defined between two column vectors as 
\begin{align}
u\cdot v=u^{T}v=\sum_{j}(u)_{j}(v)_{j}.
\end{align}
Hence, the creation operator is the scalar product between the $i$th right eigenvector of the drift matrix $\beta$ and the gradient operator. The annihilation is built from the scalar product between the left eigenvector of the drift matrix and the position and gradient operators. Similarly, in the case of the adjoint equation, the creation and annihilation operators are given by
\begin{align}
\zeta'^{+}_{i}&=\sum_{j}P^{-1}_{ij}x_{j}-\sum_{j}(P^{-1}\Sigma_{\infty})_{ij}\partial_{x_{j}}=w_{i}\cdot \underline{x}-(\Sigma_{\infty}w_{i})\cdot \underline{\partial}_{x},\\
-\zeta^{+}_{i}&=\sum_{j}P^{T}_{ij}\partial_{x_{j}}=e_{i}\cdot \underline{\partial}_{x}. 
\end{align}
Hence, we recover exactly the same raising and lowering operators as in \cite{Leen2016}, up to the normalization that differs due to conventions.
\section{Further details on the vectorization procedure \label{vectorization}}
Let us detail the representation of the operator in the vectorized formalism. The superoperator is a tensor product of operators as explained in the main text $|A\rho B \rangle=(A \otimes B^{T})|\rho\rrangle$. As a further example, consider now the  Liouville dissipator 
\begin{align}
\mathcal{D}[a](\rho)&=a\rho a^{\dagger}-\frac{1}{2}a^{\dagger}a\rho-\frac{1}{2}\rho a^{\dagger}a.
\end{align}
Its vectorized representation is given by
\begin{align}
|\mathcal{D}[a](\rho)\rrangle&=\left[a\otimes (a^{\dagger})^{T}-\frac{1}{2}a^{\dagger}a\otimes \mathbbm{1}-\frac{1}{2}\mathbbm{1}\otimes (a^{\dagger}a)^{T}\right]|\rho\rrangle.
\end{align}
However, because the matrix representation of the annihilation operator is real, $a^{\dagger}=a^{T}$ in the matrix representation given by Eq.~\eqref{rep_annihilation}. This leads to 
\begin{align}
|\mathcal{D}[a](\rho)\rrangle&=\left[a\otimes a-\frac{1}{2}a^{\dagger}a\otimes \mathbbm{1}-\frac{1}{2}\mathbbm{1}\otimes a^{\dagger}a\right]|\rho\rrangle.
\end{align}
The idea is now to convert this expression in terms of the ${\bf c},{\bf c'}$ operators, to make the dissipator quadratic in terms of these operators. Let us write
\begin{align}
{\bf c}'_{0}|\rho\rrangle&=(\mathbbm{1}\otimes a^{\dagger})|\rho\rrangle,\\
{\bf c}'_{1}|\rho\rrangle&=(a^{\dagger}\otimes\mathbbm{1})|\rho\rrangle,\\
{\bf c}_{0}|\rho\rrangle&=(-a^{\dagger}\otimes\mathbbm{1}+\mathbbm{1}\otimes a)|\rho\rrangle,\\
{\bf c}_{1}|\rho\rrangle&=(a\otimes \mathbbm{1}-\mathbbm{1}\otimes a^{\dagger})|\rho\rrangle.
\end{align}
Hence, after calculations
\begin{align}
|\mathcal{D}[a](\rho)\rrangle&=\left[{\bf c}_{0}{\bf c}_{1}+\frac{1}{2}({\bf c}_{0}{\bf c}'_{0}+{\bf c}'_{1}{\bf c}_{1})\right]\left|\rho\right\rrangle.
\end{align}
The dissipator is now a quadratic form of the operators ${\bf c}_{0},{\bf c}_{1},{\bf c}'_{0},{\bf c}'_{1}$. Further details on the vectorized formalism can be found in \cite{Gyamfi_2020}. Note also that for two matrices $A$ and $B$ the Hermitian conjugate of the tensor product is the tensor product of the Hermitian conjugate, so that
\begin{align}
(A\otimes B)^{\dagger}&=A^{\dagger}\otimes B^{\dagger}.
\end{align}
In particular, if $A$ and $B$ are real, this is simply the transpose operation. Consider the example 
\begin{align}
{\bf a}'_{0,n}&=a^{\dagger}_{n}\otimes\mathbbm{1}-\mathbbm{1}\otimes a_{n},\\
{\bf a}'_{1,n}&=a_{n}\otimes \mathbbm{1}-\mathbbm{1}\otimes a^{\dagger}_{n}.
\end{align}
As a consequence, 
\begin{align}
({\bf a}'_{0,n})^{\dagger}&=\left(a^{\dagger}_{n}\otimes\mathbbm{1}-\mathbbm{1}\otimes a_{n}\right)^{\dagger}\nonumber\\
&=a_{n}\otimes \mathbbm{1}-\mathbbm{1}\otimes a^{\dagger}_{n}={\bf a}'_{1,n}.
\end{align}
We also have
\begin{align}
{\bf c}^{\dagger}_{1,n}&=a^{\dagger}_{n}\otimes \mathbbm{1}-\mathbbm{1}\otimes a_{n}=-{\bf c}_{0,n},\\
{\bf c}^{\dagger}_{0,n}&=-a_{n}\otimes\mathbbm{1}+\mathbbm{1}\otimes a^{\dagger}_{n}=-{\bf c}_{1,n}.
\end{align}
\section{Details on the equivalence table for the $Q$ representation  \label{details_projection_table}}
Using the properties of the vectorization procedure detailed in Appendix~\ref{vectorization}, one can conjugate the equations in the main text ~\eqref{eq_c}
\begin{align}
\left({\bf c}_{0,n}|\underline{\eta}_{a}\rrangle\right)^{\dagger}&=\llangle\underline{\eta}_{a}|{\bf c}^{\dagger}_{0,n}=-\llangle\underline{\eta}_{a}|{\bf c}_{1,n}=\eta_{n}\llangle \underline{\eta}_{a}|.
\end{align}
Similarly, 
\begin{align}
\llangle \underline{\eta}_{a}|{\bf c}_{0,n}&=\eta^{*}_{n}\llangle \underline{\eta}_{a}|.
\end{align}
This gives the values for the characteristic function in Table~\ref{table1}. Using the properties of the Fourier transform Eq.~\eqref{FT1}, one obtains the equivalent equation for the $Q$ representation. This gives the two first rows of the Table~\ref{table1}. Now, using Eq.~\eqref{eq_c1} one obtains the projection to the $Q$ representation for ${\bf c}'_{0,n}$ and ${\bf c}'_{1,n}$, and using the Fourier transform Eq.~\eqref{FT2}, one obtains the corresponding equations for the characteristic function. As an example of application, let us consider the non trivial case
\begin{align}
\llangle \alpha_{n}|{\bf c}_{0,n}{\bf c}'_{0,n}|\rho\rrangle&=\int \frac{d^{2}\eta_{n}}{\pi}e^{\eta^{*}_{n}\alpha_{n}}e^{-\eta_{n}\alpha^{*}_{n}}\llangle \eta_{a,n}|{\bf c}_{0,n}{\bf c}'_{0,n}|\rho\rrangle\nonumber\\
&=\int \frac{d^{2}\eta_{n}}{\pi}e^{\eta^{*}_{n}\alpha_{n}}e^{-\eta_{n}\alpha^{*}_{n}}\eta^{*}_{n}\llangle \eta_{a,n}|{\bf c}'_{0,n}|\rho\rrangle\nonumber\\
&=\partial_{\alpha_{n}}\llangle \alpha_{n}|{\bf c}'_{0,n}|\rho\rrangle\nonumber\\
&=\partial_{\alpha_{n}}\alpha_{n}\llangle \alpha_{n}|\rho\rrangle.\label{way1}
\end{align}
As a further check, let us directly compute $\llangle \alpha_{n}|{\bf c}_{0,n}{\bf c}'_{0,n}|\rho\rrangle$
\begin{align}
&=-{\rm Tr}\left(|\alpha_{n}\rangle\langle \alpha_{n}|[a^{\dagger}_{n},\rho a_{n}]\right)\nonumber\\
&=-\langle \alpha_{n}|a^{\dagger}_{n}\rho a|\alpha_{n}\rangle+\langle \alpha_{n}|\rho a^{\dagger}a|\alpha_{n}\rangle+\langle \alpha_{n}|\rho|\alpha_{n}\rangle\nonumber\\
&=-(|\alpha_{n}|^{2}+1)\llangle \alpha_{n}|\rho\rrangle+\left(|\alpha_{n}|^{2}+\alpha_{n}\partial_{\alpha_{n}}\right)\llangle \alpha_{n}|\rho\rrangle\nonumber\\
&=\left(\alpha_{n}\partial_{\alpha_{n}}+1\right)\llangle \alpha_{n}|\rho\rrangle\nonumber\\
&=\partial_{\alpha_{n}}\alpha_{n}\llangle \alpha_{n}|\rho\rrangle. \label{way2}
\end{align}
where we used 
\begin{align}
\rho a^{\dagger}_{n}a_{n}&=a^{\dagger}_{n}\rho a_{n}-[a^{\dagger}_{n},\rho]a_{n},
\end{align}
and the property that for any function $f$ that can be expanded in series of $a$ and $a^{\dagger}$ \cite{Walls2008quantum}
\begin{align}
[a^{\dagger},f]&=\frac{\partial f}{\partial a}.
\end{align}
Hence, we verified that our method applied in Eq.~\eqref{way1} gives the same result as the computation from another method Eq.~\eqref{way2}, given in \cite{Walls2008quantum}.
\section{Details on the equivalence table for the $P$ representation \label{P_rep_table}}
In this section we provide more details on the equivalence table, Table~\ref{table}. One simple way to obtain the mapping to the $P$ representation is to use the operator correspondences\cite{Walls2008quantum}
\begin{align}
a_{n}\rho &\leftrightarrow \alpha_{n}P(\underline{\alpha}),\\
a^{\dagger}_{n}\rho &\leftrightarrow \left(\alpha^{*}_{n}-\partial_{\alpha_{n}}\right)P(\underline{\alpha}),\\
\rho a_{n} &\leftrightarrow \left(\alpha_{n}-\partial_{\alpha^{*}_{n}}\right)P(\underline{\alpha}),\\
\rho a^{\dagger}_{n}&\leftrightarrow \alpha^{*}_{n}P(\underline{\alpha}).
\end{align}
Applying this correspondence to the superoperators ${\bf a}_{0,n}, {\bf a}_{1,n}, {\bf a}'_{0,n}, {\bf a}'_{1,n}$, one finds the equivalence table, Table~\ref{table}. Another approach is to use the properties of the Fourier transform. To determine the table of equivalence for the characteristic function, we use the Fourier transform relation Eq.~\eqref{eq:fourier_transform}, which is not modified for the ordered characteristic function. By direct complex conjugation of Eq.~\eqref{eq:prosen_characteristic}
\begin{align}
\llangle \underline{\eta}_{o}|{\bf a}'_{1,n}&=-\eta_{n}\llangle \underline{\eta}_{o}|,\ & \llangle \underline{\eta}_{o}|{\bf a}'_{1,n}&=-\eta^{*}_{n}\llangle \underline{\eta}_{o}|.
\end{align}
Furthermore, from the property of the Fourier transform Eq.~\eqref{def_p_rep}
\begin{align}
&\int d^{2}\alpha_{n}\left[\alpha_{n}P(\underline{\alpha})\right]e^{-\eta^{*}_{n}\alpha_{n}}e^{\eta_{n}{\alpha}^{*}_{n}}\nonumber\\
&=-\partial_{\eta^{*}_{n}}\int d^{2}\alpha_{n}P(\underline{\alpha})e^{-\eta^{*}_{n}\alpha_{n}}e^{\eta_{n}{\alpha}^{*}_{n}}\nonumber\\
&=-\partial_{\eta^{*}_{n}}\mathcal{X}_{o}(\eta,\eta^{*}).
\end{align}
Similarly,
\begin{align}
&\int d^{2}\alpha_{n}\left[\alpha^{*}_{n}P(\underline{\alpha})\right]e^{-\eta^{*}_{n}\alpha_{n}}e^{\eta_{n}{\alpha}^{*}_{n}}\nonumber\\
&=\partial_{\eta_{n}}\int d^{2}\alpha_{n}P(\underline{\alpha})e^{-\eta^{*}_{n}\alpha_{n}}e^{\eta_{n}{\alpha}^{*}_{n}}\nonumber\\
&=\partial_{\eta_{n}}\mathcal{X}_{o}(\eta,\eta^{*}).
\end{align}
From this equation one can complete Table~\ref{table}.
\section{Consistency of our result with the literature \label{consistency_check}}
In this section, we verify that our result for the steady-state $Q$ representation recovers known results in the literature. Consider the parametrization of the coefficients of the master equation~\eqref{eq:ho_master equation}
\begin{align}
\Gamma_{\downarrow}&=2\kappa(N+1),\\
\Gamma_{\uparrow}&=2\kappa N,\\
c&=-2\kappa M,
\end{align}
with $2\kappa=\gamma_{\downarrow}-\gamma_{\uparrow}$, $M\in \mathbbm{R}$, and
\begin{align}
N=\frac{\gamma_{\uparrow}}{\gamma_{\downarrow}-\gamma_{\uparrow}}\cosh(2r)+\sinh^{2}(r).
\end{align}
Hence, taking the thermal distribution parametrization $\gamma_{\downarrow}=2\kappa(\bar{n}+1)$ and $\gamma_{\uparrow}=2\kappa\bar{n}$, this simplifies to $N=\bar{n}\cosh(2r)+\sinh^{2}(r)$. An ideally squeezed reservoir verifies the condition $M^{2}=N(N+1)$, and a nonideally correlated reservoir $M^{2}<N(N+1)$ \cite{Caves1982}. It was demonstrated in \cite{Kim1993} that for $\omega=0$, the steady-state $Q$ representation is given by 
\begin{align}
\scalebox{1}{$Q(\alpha,\infty)=\frac{1}{\pi\sqrt{(1+N)^{2}-M^{2}}}\exp\left(-\frac{\alpha^{2}_{r}}{1+N+M}-\frac{\alpha^{2}_{i}}{1+N-M}\right),$} \label{eq_reference}
\end{align}
with notation $\alpha=\alpha_{r}+i\alpha_{i}$. One can write the elements of the covariance matrix
\begin{align}
(\Sigma_{\alpha}(\infty))_{11}&=\frac{2\kappa M}{2\kappa}=M,\\
(\Sigma_{\alpha}(\infty))_{12}&=\frac{\Gamma_{\downarrow}}{\Gamma_{\downarrow}-\Gamma_{\uparrow}}=\frac{2\kappa(N+1)}{2\kappa}=(N+1),\\
(\Sigma_{\alpha}(\infty))_{22}&=\frac{2\kappa M}{2\kappa}=M.
\end{align}
Hence, the inverse covariance matrix is given by
\begin{align}
[\Sigma_{\alpha}(\infty)]^{-1}&=\frac{1}{M^{2}-(1+N)^{2}}\begin{pmatrix} M & -(1+N)\\
-(1+N) & M
\end{pmatrix}.
\end{align}
This leads to 
\begin{align}
\scalebox{1}{$\frac{1}{2}(
\alpha\
\alpha^{*})
[\Sigma_{\alpha}(\infty)]^{-1}\begin{pmatrix}
\alpha\\
\alpha^{*}
\end{pmatrix}
=\frac{\alpha^{2}_{r}}{M+1+N}+\frac{\alpha^{2}_{i}}{1+N-M}$}.
\end{align}
This recovers the exponent of Eq.~\eqref{eq_reference}.
\section{Conventions for the definition of the Wigner function and its characteristic function \label{conventions}}
Consider the rescaling $\alpha \to \sqrt{2}\alpha$ and $\eta\to \eta/\sqrt{2}$ in the definitions of the Wigner function~\eqref{def_wigner} and characteristic function~\eqref{def_charact}. Hence, the new characteristic function reads
\begin{align}
\mathcal{X}_{s}(\eta)&\equiv{\rm Tr}\left(\hat{D}^{\dagger}(\eta)\hat{\rho}\right),
\end{align}
with $\hat{D}(\alpha)\equiv e^{\alpha \hat{a}^{\dagger}-\alpha^{*}\hat{a}}$. The Wigner function and the characteristic function are related through the Fourier transform
\begin{align}
W(\alpha)&=\int \frac{ d^{2}\eta}{\pi}e^{\eta \alpha^{*}-\eta^{*}\alpha}\mathcal{X}_{w}(\eta),\\
\mathcal{X}_{s}(\eta)&=\int \frac{d^{2}\alpha}{\pi}e^{\eta^{*}\alpha-\eta\alpha^{*}}W(\alpha).
\end{align}
The product of two displacements operators remains a displacement operator
\begin{align}
\hat{D}(\beta)&=e^{(-\alpha\beta^{*}+\alpha^{*}\beta)/2}\hat{D}(\alpha)\hat{D}(\beta-\alpha).
\end{align}
After integration, the Wigner function can be seen as the expectation value of the parity operator \cite{Royer1977,Bishop1994}
\begin{align}
W(\alpha)&={\rm Tr}\left(\int \frac{d^{2}\eta}{\pi}e^{\eta \alpha^{*}-\eta^{*}\alpha}D^{\dagger}(\eta)\hat{\rho}\right)\nonumber\\
&={\rm Tr}\left(\int \frac{d^{2}\eta}{\pi}e^{\eta \alpha^{*}-\eta^{*}\alpha}D(-\eta)\hat{\rho}\right)\nonumber\\
&={\rm Tr}\left(\int \frac{d^{2}\eta}{\pi}e^{\eta \alpha^{*}-\eta^{*}\alpha}e^{(\beta\eta^{*}-\beta^{*}\eta)/2}\hat{D}(\beta)\hat{D}(-\eta-\beta)\hat{\rho}\right).
\end{align}
Hence, choosing $\beta=2\alpha$, one obtains
\begin{align}
W(\alpha)&={\rm Tr}\left(\hat{D}(2\alpha)\int \frac{d^{2}\eta}{\pi}\hat{D}(-\eta-2\alpha)\hat{\rho}\right)\nonumber\\
&={\rm Tr}\left(\hat{D}(2\alpha)\int \frac{d^{2}\eta}{\pi}\hat{D}(-\eta-2\alpha)\hat{\rho}\right).
\end{align}
Using the property of the integration of the displacement operator, demonstrated in \cite{Bishop1994}
\begin{align}
\int d^{2}\eta\hat{D}(\eta)&=2\pi e^{i\pi a^{\dagger}a}.
\end{align}
This leads to
\begin{align}
W(\alpha)&=2{\rm Tr}\left(\hat{D}(2\alpha)e^{i\pi a^{\dagger}a}\hat{\rho}\right)\nonumber\\
&=2{\rm Tr}\left(e^{i\pi a^{\dagger}a}\hat{D}^{\dagger}(2\alpha)\hat{\rho}\right),
\end{align}
where we use the property of the parity operator
$a e^{i\pi a^{\dagger}a}=-e^{i\pi a^{\dagger}a}a$, leading to $e^{-i\pi a^{\dagger}a}D(2\alpha)e^{i\pi a^{\dagger}a}=D(-2\alpha)$.

\bibliography{phase_space}
\end{document}